\documentclass[a4paper,11pt]{article}

\pdfoutput=1

\newcommand{\den}{\rho(\vec{x},\theta,t)}
\newcommand{\denbar}{\bar{\rho}(\vec{x},\theta,t)}
\newcommand{\denp}{\rho(\vec{x},\theta^\prime,t)}
\newcommand{\denpbar}{\bar{\rho}(\vec{x},\theta^\prime,t)}

\usepackage{jcappub}
\usepackage{amsmath}
\usepackage{amsfonts}
\usepackage{amssymb}
\usepackage{mathtools}
\usepackage{graphics}
\usepackage{citesort}
\usepackage{graphicx}
\usepackage{url}
\usepackage{xcolor}
\usepackage{cancel}
\usepackage[normalem]{ulem}
\usepackage[utf8]{inputenc}
\usepackage{subcaption}
\usepackage{hyperref}

\usepackage{placeins}

\title{Multi-Dimensional Solution of Fast Neutrino Conversions in Binary Neutron Star Merger Remnants}

\author{Ian Padilla-Gay,}
\author{Shashank Shalgar,}
\author{and Irene Tamborra}
\affiliation{Niels Bohr International Academy and DARK, Niels Bohr Institute, University of Copenhagen, Blegdamsvej 17, 2100, Copenhagen, Denmark}

\emailAdd{ian.padilla@nbi.ku.dk}
\emailAdd{shashank.shalgar@nbi.ku.dk}
\emailAdd{tamborra@nbi.ku.dk}

\abstract{Fast pairwise conversions of neutrinos are predicted to be ubiquitous in neutron star merger remnants with potentially major implications on the nucleosynthesis of the elements heavier than iron. We present the first sophisticated numerical solution of the neutrino flavor evolution above the remnant disk within a (2+1+1) dimensional setup: two spatial coordinates, one angular variable, and time. We look for a steady-state flavor configuration above the remnant disk. Albeit the linear stability analysis predicts flavor instabilities at any location above the remnant disk, our simulations in the non-linear regime show that fast pairwise conversions lead to minimal neutrino mixing ($<1\%$); flavor equilibration is never achieved in our models. Importantly, fast neutrino conversions are more prominent within localized regions near the edges of the (anti)neutrino decoupling surfaces and almost negligible in the polar region of the remnant. Our findings on the role of fast pairwise conversions should be interpreted with caution because of the approximations intrinsic to our setup and advocate for further work within a more realistic  framework.
}

\begin{document}
\maketitle
\flushbottom

\section{Introduction}
\label{sec:intro}
The coalescence of a neutron star (NS) with another NS or a black hole (BH) leads to the birth of a compact binary merger. Gravitational waves (GW) from a binary neutron star merger have been detected by the LIGO and Virgo Collaborations, the GW170817 event, together with the multi-wavelength electromagnetic counterpart~\cite{TheLIGOScientific:2017qsa,Monitor:2017mdv,GBM:2017lvd}. The multi-messenger detection of GW170817 has confirmed theoretical predictions according to which compact binary mergers are the precursors of short gamma-ray bursts (sGRBs), one of the main factories where the elements heavier than iron are synthesized---through the rapid neutron-capture process ($r$-process)---and power kilonovae (electromagnetic transients bright in the optical and infrared wavebands)~\cite{1974ApJ192L.145L,Eichler:1989ve,Li:1998bw,Kulkarni:2005jw,Metzger:2010sy,Metzger:2016pju}.

The GW170817 observation has shed light on the poorly explored physics of NS mergers. However, a robust theoretical understanding of the physics of these objects is still lacking and three-dimensional general-relativistic magnetohydrodynamical simulations with detailed neutrino transport are not yet available. In particular, the role of neutrinos is especially unclear despite the fact that a copious amount of neutrinos is produced in the coalescence. Neutrinos should affect the cooling of the merger remnant, as well as the overall ejecta composition, and contribute to power sGRBs~\cite{Wanajo:2014wha,Perego:2014fma,Fernandez:2015use,Sekiguchi:2015dma,Radice:2016dwd,Miller:2019dpt,Eichler:1989ve,Woosley:1993wj,Ruffert:1998qg,2011MNRAS.410.2302Z,Just:2015dba,Foucart:2020qjb}. 

A crucial ingredient possibly affecting the neutrino reaction rates and energy deposition is the neutrino flavor conversion physics, currently neglected in most of the literature on the subject. 
Besides the ordinary interactions of neutrinos with matter~\cite{Wolfenstein:1977ue, Mikheev:1986gs}, in compact binary mergers, the neutrino density is so high that $\nu$--$\nu$ interactions cannot be neglected, similarly to the case of core-collapse supernovae~\cite{Mirizzi:2015eza,Duan:2010bg,Chakraborty:2016yeg}. A characteristic feature of compact binary mergers is the excess of $\bar\nu_e$ over $\nu_e$ due to the overall protonization of the merger remnant~\cite{Ruffert:1996by,Foucart:2015vpa,Perego:2014fma}. As a consequence, a matter-neutrino resonance can occur as the matter potential cancels the $\nu$--$\nu$ potential~\cite{Malkus:2012ts,Malkus:2014iqa,Wu:2015fga,Zhu:2016mwa,Frensel:2016fge,Tian:2017xbr,Shalgar:2017pzd}.

In addition to the matter-neutrino resonance, $\nu$--$\nu$ interactions can be responsible for the development of fast pairwise conversions~\cite{Sawyer:2008zs,Sawyer:2015dsa,Izaguirre:2016gsx}. The latter can be triggered by the occurrence of 
electron lepton number (ELN) crossings in the neutrino angular distributions and could lead to flavor conversions on a time scale  $G_{\textrm{F}}|n_{\nu_e}-n_{\bar{\nu}_e}|^{-1}$, where $G_{\textrm{F}}$ is the Fermi constant and $n_{\nu_e}$ ($n_{\bar\nu_e}$) is the local number density of $\nu_e$ ($\bar\nu_e$). Reference~\cite{Wu:2017qpc} pointed out that fast pairwise conversions could be ubiquitous above the remnant disk because of the accretion torus geometry and the natural protonization of the remnant leading to an excess of $\bar\nu_e$ over $\nu_e$.

Whether flavor equipartition is achieved as a consequence of fast pairwise conversions is a subject of intense debate, also in the context of core-collapse supernovae~\cite{Shalgar:2019kzy,Johns:2019izj,Shalgar:2019qwg,Abbar:2017pkh,Shalgar:2020xns,Capozzi:2020kge,Abbar:2018beu,Capozzi:2018clo,Bhattacharyya:2020dhu}. 
If fast pairwise conversions lead to flavor equilibration in compact binary mergers, the nucleosynthesis of the heavy elements in the neutrino-driven wind can be drastically affected, and the fraction of lanthanides boosted with major implications for the kilonova observations~\cite{Wu:2017drk}. The possible consequences of fast pairwise conversions on the physics of compact mergers justify a modeling of the flavor conversion physics that goes beyond the predictions of the linear stability analysis~\cite{Banerjee:2011fj,Raffelt:2013rqa,Izaguirre:2016gsx}.

 Building on Ref.~\cite{Shalgar:2019qwg}, we present the first sophisticated modeling of fast pairwise conversions in the non-linear regime above the disk of merger remnants. We rely on a (2+1+1) dimensional setup: we track the neutrino flavor evolution in two spatial coordinates, one angular variable, and time. We solve the equations of motion of (anti)neutrinos in the absence of collisions and aim to investigate the flavor evolution of the neutrino-dense gas above the disk of the remnant, searching for a steady-state configuration of flavor. 
Our goal is to identify the location and extent of regions with significant flavor conversion.

This work is organized as follows. In Section~\ref{sec:setup}, we introduce the 2D box configuration that we adopt to model the neutrino emission and propagation above the merger remnant disk. The neutrino equations of motion and the semi-analytical tools to explore the eventual occurrence of flavor instabilities in the context of fast pairwise conversions are introduced in Sec.~\ref{sec:fast}. In Sec.~\ref{sec:nsm}, we present our findings on fast pairwise conversions above the massive NS remnant disk in the non-linear regime; we also explore how the steady-state flavor configuration is affected by variations of the input model parameters. In Sec.~\ref{sec:BHdisk}, we investigate the flavor oscillation physics above a BH remnant disk. Finally, our conclusions are presented in Sec.~\ref{sec:conc}. The routine adopted to take into account the effects of neutrino advection in the presence of flavor conversions is outlined in Appendix~\ref{appendix:algorithm}. Appendix~\ref{appendix:spatial} instead provides details on the convergence of our results for the adopted spatial resolution.

\section{Merger remnant disk setup}
\label{sec:setup}

\begin{figure}
\centering
\includegraphics[width=0.60\textwidth]{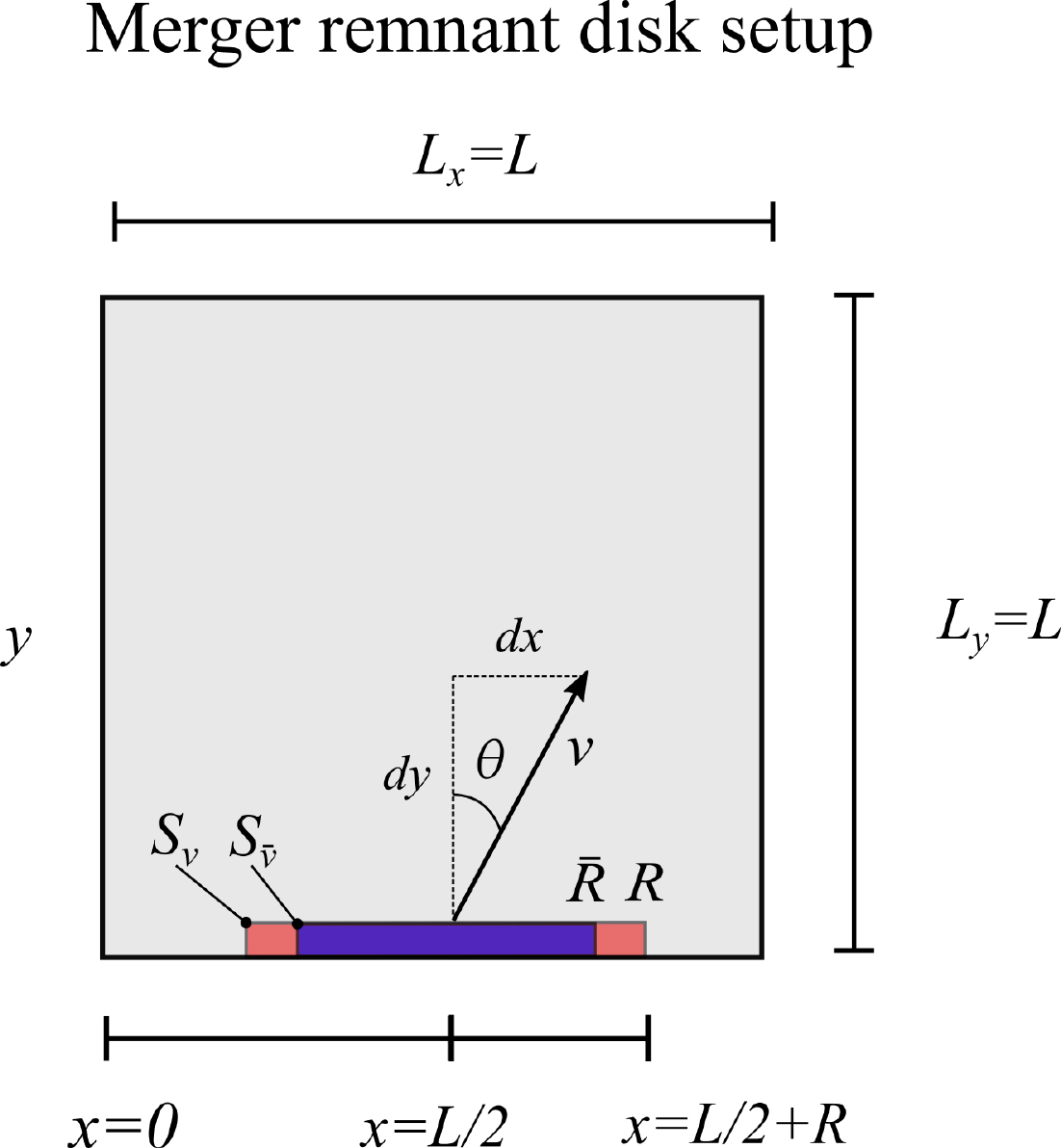}
\caption{Schematic representation of our merger remnant setup in a 2D box of size $L_x=L_y=L$. The neutrino source  ($\mathcal{S}_{\nu}$, in red) and the antineutrino one  ($\mathcal{S}_{\bar{\nu}}$, in blue) have widths $2 R$ and $2 \bar{R}$ and are centered on $(x,y)=(L/2,0)$, respectively. The (anti)neutrino sources emit $\nu_e$ and $\bar{\nu}_e$ in the forward direction only i.e. $\theta \in (-\pi/2,\pi/2)$.
}
\label{fig:0}
\end{figure}

Given the numerical challenges involved in the modeling of the flavor conversion physics within a realistic astrophysical framework, we here focus on a simpler toy model inspired by the ones adopted in Refs.~\cite{Wu:2017qpc,Shalgar:2019qwg}. We model the neutrino emission above the remnant disk in a 2D box with width $L_{x}$ and height $L_{y}$, with $L_x = L_y \equiv L = 80$ km, as sketched in Fig.~\ref{fig:0}. Although this is a small patch of the overall region above the merger remnant, it is large enough to explore the development and evolution of fast pairwise conversions. First, we model a NS-disk remnant; our findings are extended to the case of a BH-disk remnant in Sec.~\ref{sec:BHdisk}. At the bottom edge of the grid ($y = 0$) in Fig.~\ref{fig:0}, we locate a thin neutrino source, $\mathcal{S}_\nu$, of length $R = L/4 = 20$~km, which represents the $\nu_e$ neutrinosphere. Similarly, we consider a source of $\bar\nu_e$, $\mathcal{S}_{\bar{\nu}}$, of length $\bar{R} = 75\% R$. 
The neutrino and antineutrino emission surfaces are centered on $(x = L/2, y = 0)$.
 The neutrino source $\mathcal{S}_{\nu}$ is such that $x \in [L/2 - R, L/2 + R]$ and similarly for $\mathcal{S}_{\bar{\nu}}$ with the replacement $R \rightarrow \bar{R}$. 
 
Our choice of the $\mathcal{S}_{\nu}$ size with respect to the one of $\mathcal{S}_{\bar{\nu}}$ is guided by hydrodynamical simulations of a massive NS-disk~\cite{Perego:2014fma}.  Although it is well known that the decoupling surfaces of $\nu_e$ and $\bar\nu_e$ are spatially well separated, see e.g.~\cite{Foucart:2015vpa,Wu:2017drk}, we assume that the neutrinospheres of $\nu_e$ and $\bar\nu_e$ are coincident and the decoupling occurs suddenly for the sake of simplicity. As we will discuss later, this has an impact on the formation of ELN crossings, but it does not affect the overall flavor conversion picture above the remnant disk.
 
 We also assume that non-electron flavors are generated through flavor conversions only. In the case of NS-disk remnants, a small amount of non-electron (anti)neutrinos is naturally produced in the NS-disk remnant (see, e.g., Refs.~\cite{Perego:2014fma,Ardevol-Pulpillo:2018btx}); in this case, our extreme assumption enhances the likelihood of having flavor conversions and, as we will discuss in Sec.~\ref{sec:nsm}, it does not affect our overall conclusions. Our ansatz closely mimics the BH-disk remnant case instead (see Sec.~\ref{sec:BHdisk} and, e.g., Ref.~\cite{Just:2014fka}). 
 
 As for the boundary conditions in our 2D box, we assume that, except for the edge containing the (anti)neutrino sources, the other edges of the 2D box act as sinks for (anti)neutrinos. Since the (anti)neutrinos continuously flow from the sources into the sinks, the total number density of neutrinos and antineutrinos is conserved.

We work in a two-flavor approximation, $(\nu_e,\nu_x)$, and denote with $\nu_x$ a mixture of the non-electron flavors. In order to describe the neutrino and antineutrino fields, we rely on $2 \times 2$ density matrices defined for each $(x,y)$ point in the 2D box: 
\begin{eqnarray}
\den = 
\begin{pmatrix}
\rho_{ee} & \rho_{ex}\\
\rho_{ex}^{*} & \rho_{xx} 
\end{pmatrix}\ \quad\ \mathrm{and}\ \quad\ 
\denbar =
\begin{pmatrix}
\bar{\rho}_{ee} & \bar{\rho}_{ex}\\
\bar{\rho}_{ex}^{*} & \bar{\rho}_{xx} 
\end{pmatrix}\ .
\end{eqnarray}
The diagonal terms of the density matrix encode the flavor content information and are proportional to the (anti)neutrino number densities in $(x,y)$; the off-diagonal terms are connected to the probability of flavor transitions, as we will discuss in the next section. As such, we normalize the density matrices in the following way: $\mathrm{tr}(\rho)=1$ and $\mathrm{tr}(\bar\rho)= a$. The parameter $a$ takes into account the asymmetry between neutrinos and antineutrinos, and we take $a = 2.4$~\cite{Wu:2017drk} in the numerical runs. 

The flavor conversion physics is affected by the distributions in angle and in energy of neutrinos and antineutrinos. However, since we focus on fast pairwise conversions, in the following we assume a  monoenergetic distribution of (anti)neutrinos, with average energy $\langle E_{\nu_e} \rangle = \langle E_{\bar\nu_e} \rangle \simeq 20$~MeV to mimic typical average energies in the proximity of the (anti)neutrino decoupling region; as shown in Ref.~\cite{Shalgar:2020xns}, the assumption of a  monoenergetic distribution reproduces the flavor outcome that one would obtain when an energy distribution centered on $\langle E_{\nu_e,\bar\nu_e} \rangle$ is considered. As a consequence, the neutrino distribution, for each point in the $(x,y)$ box and at the time $t$, is defined by the emission angle $\theta$ (see Fig.~\ref{fig:0}).

In order to model the physics of fast pairwise conversions, we need to take into account the (anti)neutrino angular distributions. We assume that the emission surfaces of neutrinos and antineutrinos are perfect black-bodies and (anti)neutrinos are uniformly emitted in the forward direction across the source, i.e., $\theta \in (-\pi/2,\pi/2)$ with $\theta$ measured with respect to the $y$ direction (see Fig.~\ref{fig:0}). In order to guarantee that the emitting surfaces are Lambertian and the neutrino radiance is the same along any viewing angle, we assume the angular distributions to be proportional to $\cos{\theta}$: 
\begin{eqnarray}
\rho_{ee}(\theta) &=& 
\cos{\theta} \times
\begin{cases}
1 & \mathrm{if} \ x_{0,\nu} \in \mathcal{S}_{\nu} \\
\exp{\Big(\frac{(x - L/2 \mp R)^2}{2 \sigma^2}\Big)} & \mathrm{otherwise} \ ,
\end{cases} 
\label{case1}
\\
\bar{\rho}_{ee}(\theta) &=& 
a \cos{\theta} \times
\begin{cases}
1 & \mathrm{if} \ x_{0,\bar{\nu}} \in \mathcal{S}_{\bar{\nu}} \\
\exp{\Big(\frac{(x - L/2 \mp \bar{R})^2}{2 \bar{\sigma}^2}\Big)} & \mathrm{otherwise} \ ,
\end{cases} 
\label{case2}
\end{eqnarray}
where $\sigma$, $\bar{\sigma}$ smooth the edges of $\mathcal{S}_{\nu}$ and $\mathcal{S}_{\bar{\nu}}$ and are set to $20 \% R$ and $20 \% \bar{R}$, respectively.

\begin{figure}
\centering
\includegraphics[width=0.95\textwidth]{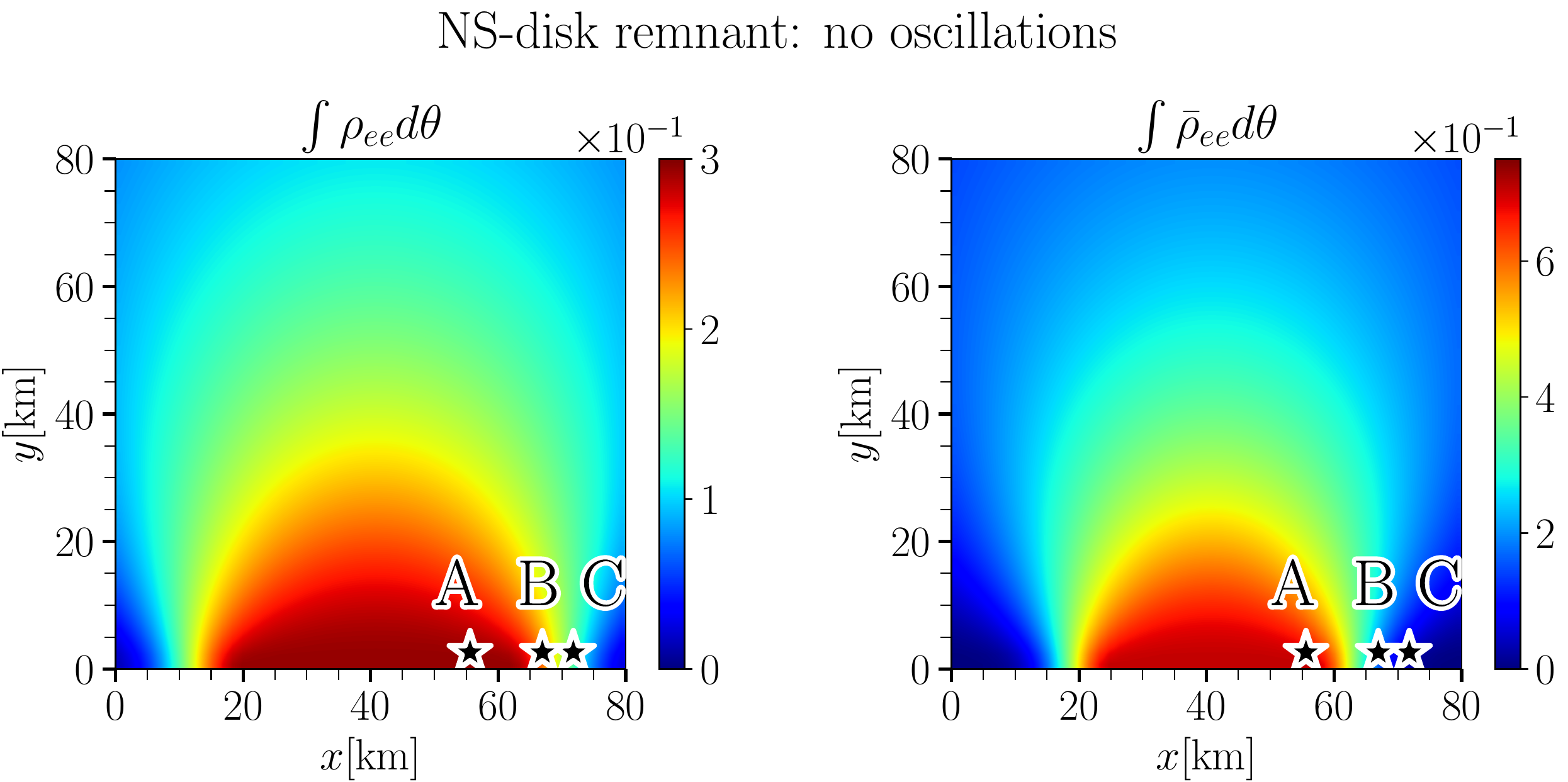}
\caption{Contour plots of the angle-integrated density matrices, $\den$ (on the left) and $\denbar$ (on the right), in the absence of flavor conversions for the NS-disk remnant configuration. This configuration can be obtained by solving Eqs.~\ref{eom1} and \ref{eom2} for $H(\theta)=\bar{H}(\theta)=0$, i.e.~the time evolution of $\den$ and $\denbar$ is completely determined by the advective operator $ \vec{v}\cdot\vec{\nabla}$ (see Sec.~\ref{sec:fast} for more details). The quantities $\int \rho_{ee}d\theta$ and $\int \bar{\rho}_{ee}d\theta$ are normalized to the maximum total particle number in the box [$\int (\rho_{ee}+\bar{\rho}_{ee}+2\rho_{xx}) d\theta$].
The coordinates of the points A, B, and C marked on the plane are: $(x, y) \simeq (56, 1)$~km, $(67, 1)$~km, and $(72, 1)$~km, respectively. The (anti)neutrino density gradually decreases as one moves away from $\mathcal{S}_{\nu}$ and $\mathcal{S}_{\bar\nu}$.
}
\label{fig:1}
\end{figure}

By projecting the neutrino and antineutrino angular distributions from the sources on any $(x,y)$ point in the 2D box, we obtain the contour plots in Fig.~\ref{fig:1} for the resultant angle-integrated density matrices of $\nu_e$ and $\bar\nu_e$, $\int d\theta \rho_{ee}(\vec{x}, \theta)$ and $\int d\theta \bar\rho_{ee}(\vec{x}, \theta)$, in the absence of flavor conversions (see also Sec.~\ref{sec:fast}) for the NS-disk remnant configuration. One can see that the neutrino density gradually decreases as one moves from $\mathcal{S}_{\nu}$ and $\mathcal{S}_{\bar\nu}$ towards the edges of the box.
\begin{figure}
\centering
\includegraphics[width=0.6\textwidth]{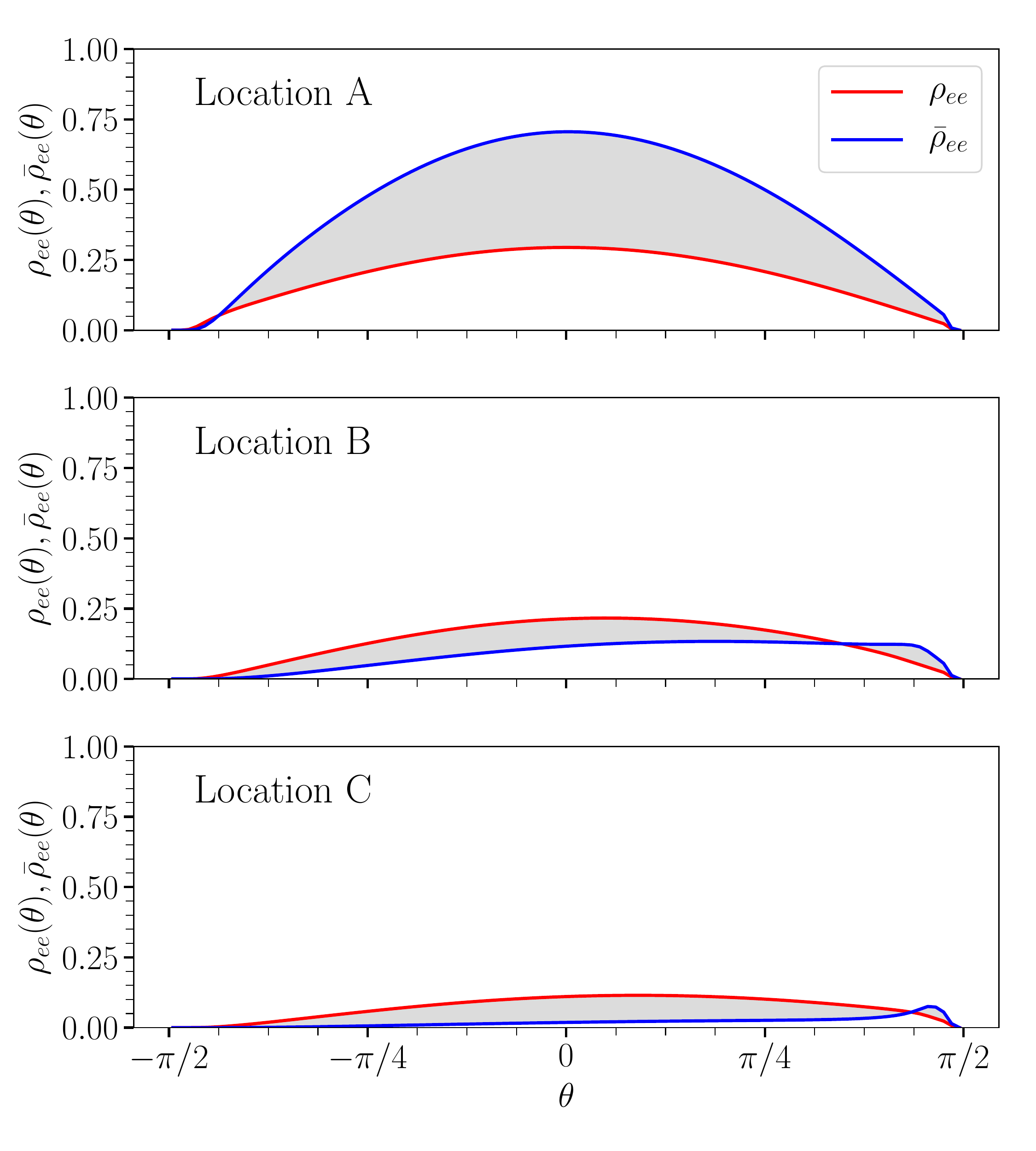}
\caption{
Angular distributions of $\rho_{ee}$ (red) and $\bar{\rho}_{ee}$ (blue) in A, B and C (see Fig.~\ref{fig:1}).  The presence of two disjoint grey areas imply the existence of ELN angular crossings. The angle-dependent density matrix elements are normalized to the maximum total particle number in $(x,y)_{\mathrm{A,B,C}}$: $(\rho_{ee}+\bar{\rho}_{ee}+2\rho_{xx})$. As one moves away from $\mathcal{S}_{\nu}$ and $\mathcal{S}_{\bar\nu}$, the width of the ELN crossings varies with implications on the flavor conversion physics.
}
\label{fig:1aa}
\end{figure}

In order to explore the variation of the (anti)neutrino angular distributions across the 2D box, Fig.~\ref{fig:1aa} displays the angular distributions of $\nu_e$ and $\bar{\nu}_e$ in the points A, B, and C highlighted in Fig.~\ref{fig:1}. The width of the ELN crossings varies as one moves away from $\mathcal{S}_{\nu}$ and $\mathcal{S}_{\bar\nu}$, with implications on the flavor conversion physics, as we discuss in Sec.~\ref{sec:fast}. 

\section{Fast pairwise neutrino flavor conversion}
\label{sec:fast}
In this section, we introduce the equations of motion adopted to track the flavor evolution above the NS-disk remnant in our 2D setup. In order to gauge the role of neutrino flavor conversions above the NS-disk remnant, we discuss the variation of the $\nu$--$\nu$ interaction strength across the 2D box and introduce the instability parameter to characterize the depth of the ELN crossings. We also adopt the linear stability analysis to compute the growth rate of the flavor instabilities in the regions with the largest instability parameter.

\subsection{Equations of motion}

The (anti)neutrino field is described through the density matrix approach introduced in Sec.~\ref{sec:setup}. 
Neglecting collisions, the flavor evolution of neutrinos and antineutrinos is described by the following set of equations of motion (EoM): 
\begin{eqnarray}
i\left(\frac{\partial}{\partial t} + \vec{v}\cdot\vec{\nabla}\right) \den
&=& [H(\theta),\den]\ ,
\label{eom1}
\\
i\left(\frac{\partial}{\partial t} + \vec{v}\cdot\vec{\nabla}\right) \denbar
&=& [\bar{H}(\theta),\denbar]\ ,
\label{eom2}
\end{eqnarray}
where the advective term, $\vec{v}\cdot\vec{\nabla}$, is proportional to the velocity of (anti)neutrinos, which we assume to be equal to the speed of light, and is tangential to the neutrino trajectory. The contour plots of the angle-integrated density matrices of $\nu_e$ and $\bar\nu_e$ in Fig.~\ref{fig:1} can be obtained by solving Eqs.~\ref{eom1} and \ref{eom2} when the right hand side of both EoMs is vanishing, i.e., (anti)neutrinos do not change their flavor. 

The neutrino Hamiltonian is 
\begin{eqnarray}
\label{eq:H}
H(\theta) = 
\frac{\omega}{2} \left(\begin{matrix}-\cos 2\theta_V & \sin 2\theta_V\\ \sin 2\theta_V & \cos 2\theta_V\end{matrix}\right)+ H_{\nu\nu}(\vec{x},\theta) \ ,
\end{eqnarray}
with the first term depending on the vacuum frequency $\omega = 0.3$~km$^{-1}$, where $\omega = \Delta m^2/2 \langle E_{\nu_e,\bar\nu_e} \rangle$, $\Delta m^2$ is the atmospheric squared mass difference and $\langle E_{\nu_e,\bar\nu_e}\rangle$ the average mean energy of $\nu_e$'s and $\bar\nu_e$'s introduced in Sec.~\ref{sec:setup}. The vacuum mixing angle is $\theta_{V}=10^{-6}$; note that we assume a very small mixing to effectively ignore the matter potential~\cite{EstebanPretel:2008ni}. The second term of the Hamiltonian is the $\nu$--$\nu$ interaction term: 
\begin{eqnarray}
\label{eq:Hnunu}
H_{\nu\nu}(\vec{x},\theta) = 
\mu(|\vec{x}|) \int d \theta^\prime \left[\denp - \denpbar \right] \left[1 - \cos(\theta - \theta^{\prime})\right]\ .
\end{eqnarray}
The potential, $\mu(|\vec{x}|)$, parametrizes the strength of neutrino-neutrino interactions for each point $(x,y)$ in the box and its functional form is defined in Sec~\ref{sec:mu}.
The Hamiltonian of antineutrinos, $\bar{H}(\theta)$, is identical to $H(\theta)$ except for the following replacement: $\omega \rightarrow -\omega$~\cite{Duan:2006an}. 
The integration over $d\theta^{\prime}$ is a consequence of our 2D setup. In a 3D box, the integration over $d\theta^{\prime}$ would be replaced by an integration over the solid angle. We have checked, however, that the integration over $d\cos\theta$ that would arise in an azimuthally symmetric 3D  system virtually gives  the same results as our 2D setup (see also Sec.~\ref{sec:mu}).

\subsection{Neutrino self-interaction potential}
\label{sec:mu}

The $\nu$--$\nu$ interaction potential varies across our 2D box, by taking into account the dilution of the (anti)neutrino gas as we move away from the sources $\mathcal{S}_{\nu}$ and $\mathcal{S}_{\bar{\nu}}$. We parametrize it as 
\begin{eqnarray}
\label{eq:mudef}
 \mu(|\vec{x}|) = \mu_0 \eta(|\vec{x}|)\ ,
\end{eqnarray} 
 where $\eta(|\vec{x}|)$ is a scaling function, and $\mu_0 = 10^5\ \mathrm{km}^{-1}$ is the $\nu$--$\nu$ interaction strength at the neutrinosphere~\cite{Wu:2017qpc}.

\begin{figure}
\centering
\includegraphics[width=0.7\textwidth]{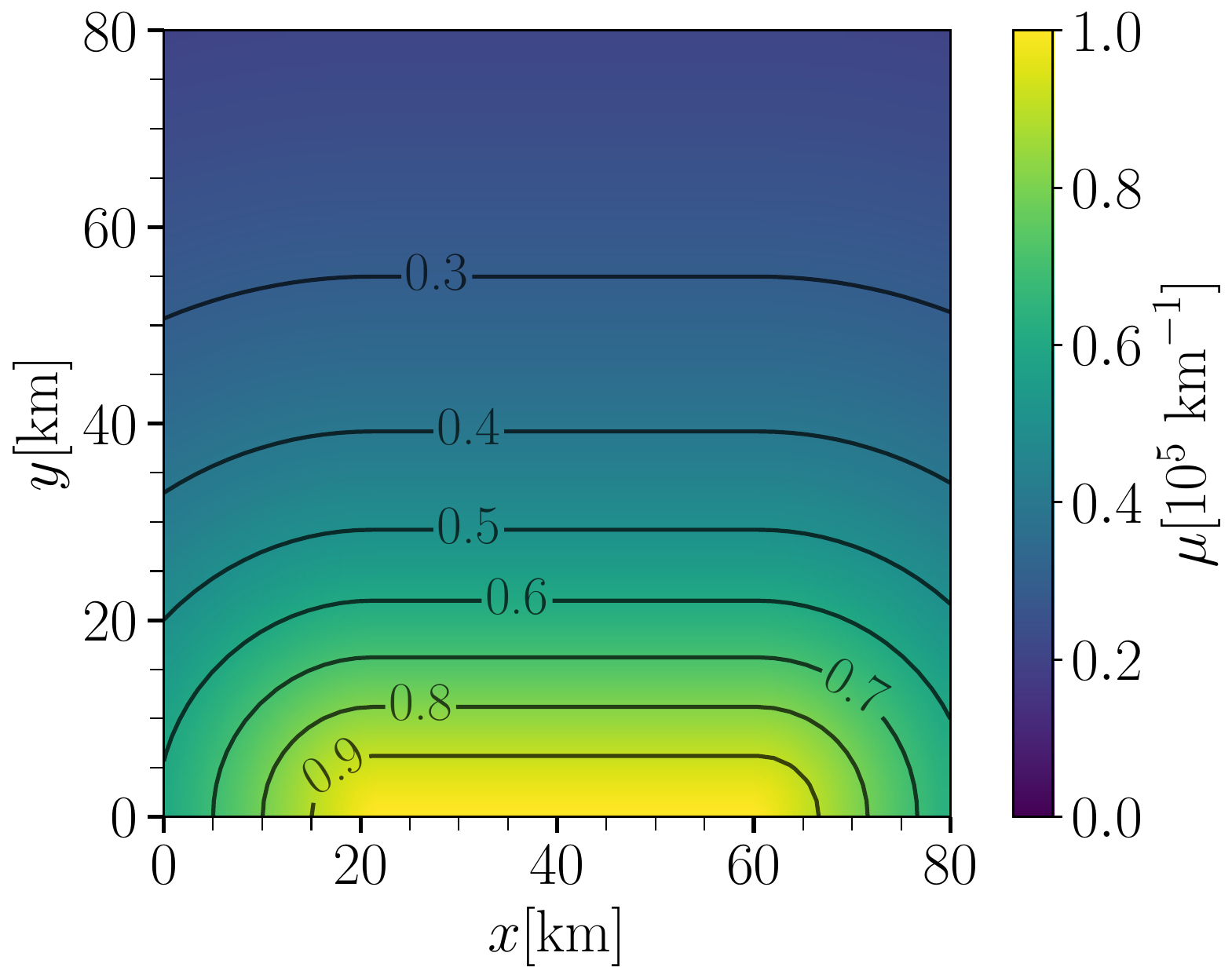}
\caption{Contour plot of the neutrino self-interaction strength, $\mu(|\vec{x}|)$, in the 2D box. The neutrino-neutrino potential is maximum in the proximity of the (anti)neutrino source and it gradually decreases as the distance from the (anti)neutrino sources increases.
}
\label{fig:30}
\end{figure}

Since  the modeling of flavor evolution in 3D is computationally challenging at present, we mimic the 3D setup by solving the EoM in 2D while taking into account the dilution of the neutrino gas in 3D. 
For an observer located at $(x,y)$, the distance $d$ above the source, $\mathcal{S}_{\nu,\bar\nu}$, can be computed as $d=dy/\cos{\theta}$, where $dy$ is the vertical displacement from the source to $(x,y)$, see Fig.~\ref{fig:0}. For observers that are not located above the source, the dilution of the flux is determined by the distance $d$.
With this convention, the scaling function $\eta$ is defined as
\begin{eqnarray}
\label{eq:geomratio}
 \eta = \Bigg( 1 - \frac{1}{\sqrt{(R/d)^2+1}} \Bigg)^2 \Bigg[ \arccos{\Bigg( \frac{1}{\sqrt{ (R/d)^2 + 1 }} \Bigg)} - \sqrt{1 - \frac{1}{(R/d)^2 + 1 } } \Bigg]^{-1} \ .
\end{eqnarray}
To better understand the role of $\eta$, let us look at one limiting case for an observer along the axis of symmetry.
When $dy \gg R$, $\eta \propto (R/dy)$, hence $H_{\nu\nu} \propto (R/dy)^4$ for a 3D bulb model, as expected~\cite{Duan:2006an}. 
Figure~\ref{fig:30} shows $\mu(|\vec{x}|)$ in our 2D box (see Eq.~\ref{eq:mudef}). At the (anti)neutrino emission surfaces, $\mu$ assumes the maximum value ($\mu_0$) and drops as a function of the distance from the source. 

\subsection{Instability parameter}
A favorable condition for the development of fast pairwise conversions is the presence of ELN crossings between the angular distributions of $\nu_e$ and $\bar\nu_e$~\cite{Izaguirre:2016gsx}. 
 To this purpose, the ``instability parameter" has been introduced in Ref.~\cite{Shalgar:2019qwg} to gauge the growth rate of flavor instabilities; the latter being dependent on the depth of the ELN crossings~\cite{Martin:2019gxb,Yi:2019hrp}:
\begin{eqnarray}
\zeta &=& \rho_{\mathrm{tot}}\frac{I_{1}I_{2}}{(I_{1}+I_{2})^{2}}\ , 
\label{zeta:def}
\end{eqnarray}
where $\rho_{\mathrm{tot}}$ is the total particle number defined as $\int[\rho_{ee}+\bar{\rho}_{ee}+2\rho_{xx}]d\theta$ and the factors $I_{1,2}$ are defined as
\begin{eqnarray}
\label{I1:def}
I_{1} = \int_{-\pi/2}^{\pi/2} \Theta \left[\rho_{ee}(\theta)-\bar{\rho}_{ee}(\theta)\right] d\theta\ \mathrm{and}\ 
I_{2} = \int_{-\pi/2}^{\pi/2} \Theta \left[\bar{\rho}_{ee}(\theta)-\rho_{ee}(\theta)\right] d\theta\ ; 
\end{eqnarray} 
the Heaviside function, $\Theta$, is vanishing for $\rho_{ee}(\theta)-\bar{\rho}_{ee}(\theta) < 0$ and otherwise equal to the identity operator. 
The instability parameter $\zeta$ vanishes when the ELN crossing is zero. The instability parameter is a useful predictor of the growth rate of the off-diagonal components of the density matrices and, therefore, of the flavor instabilities (see Sec.~3.4 of Ref.~\cite{Shalgar:2019qwg} for more details).

\begin{figure}
\centering
\includegraphics[width=0.49\textwidth]{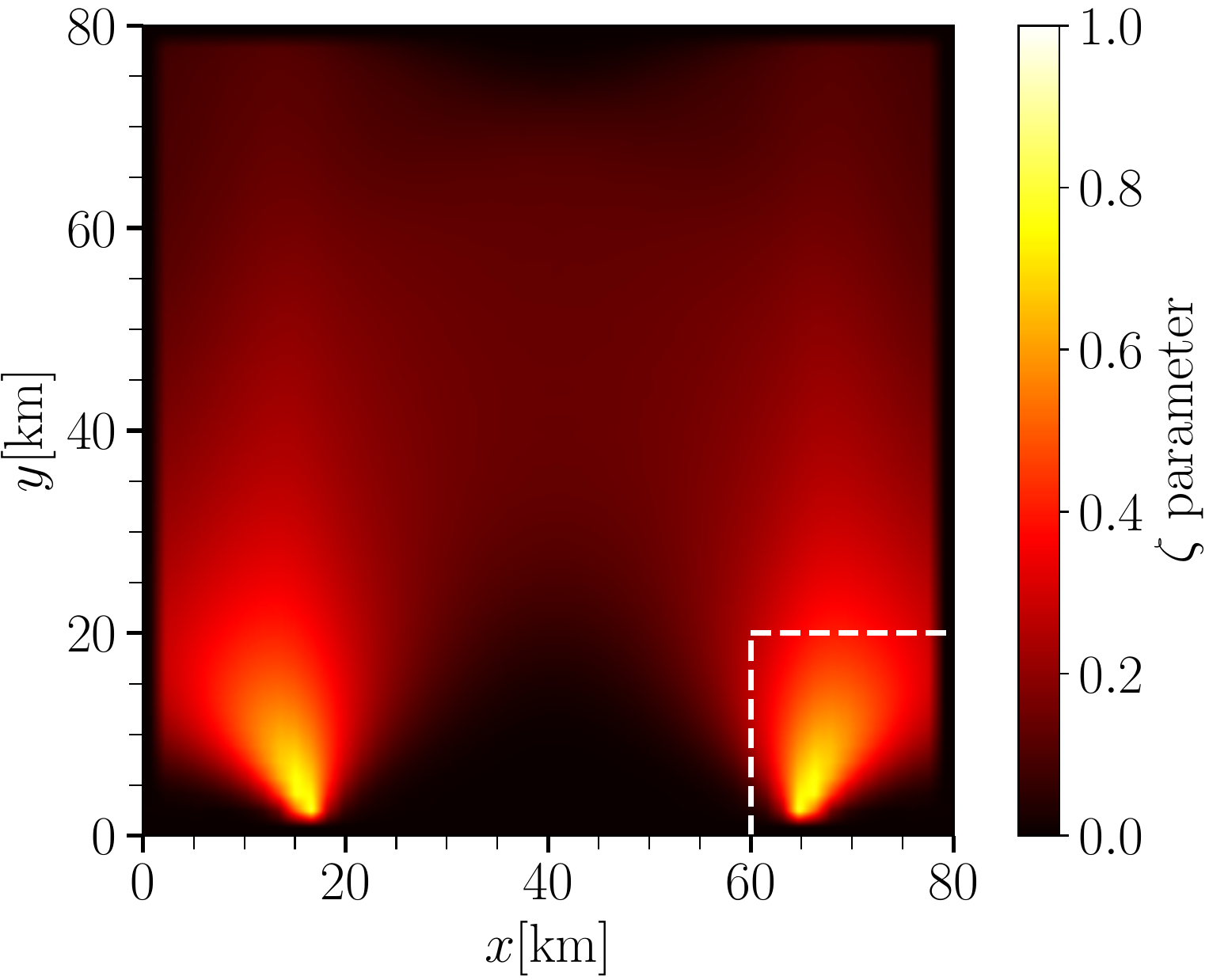}
\includegraphics[width=0.48\textwidth]{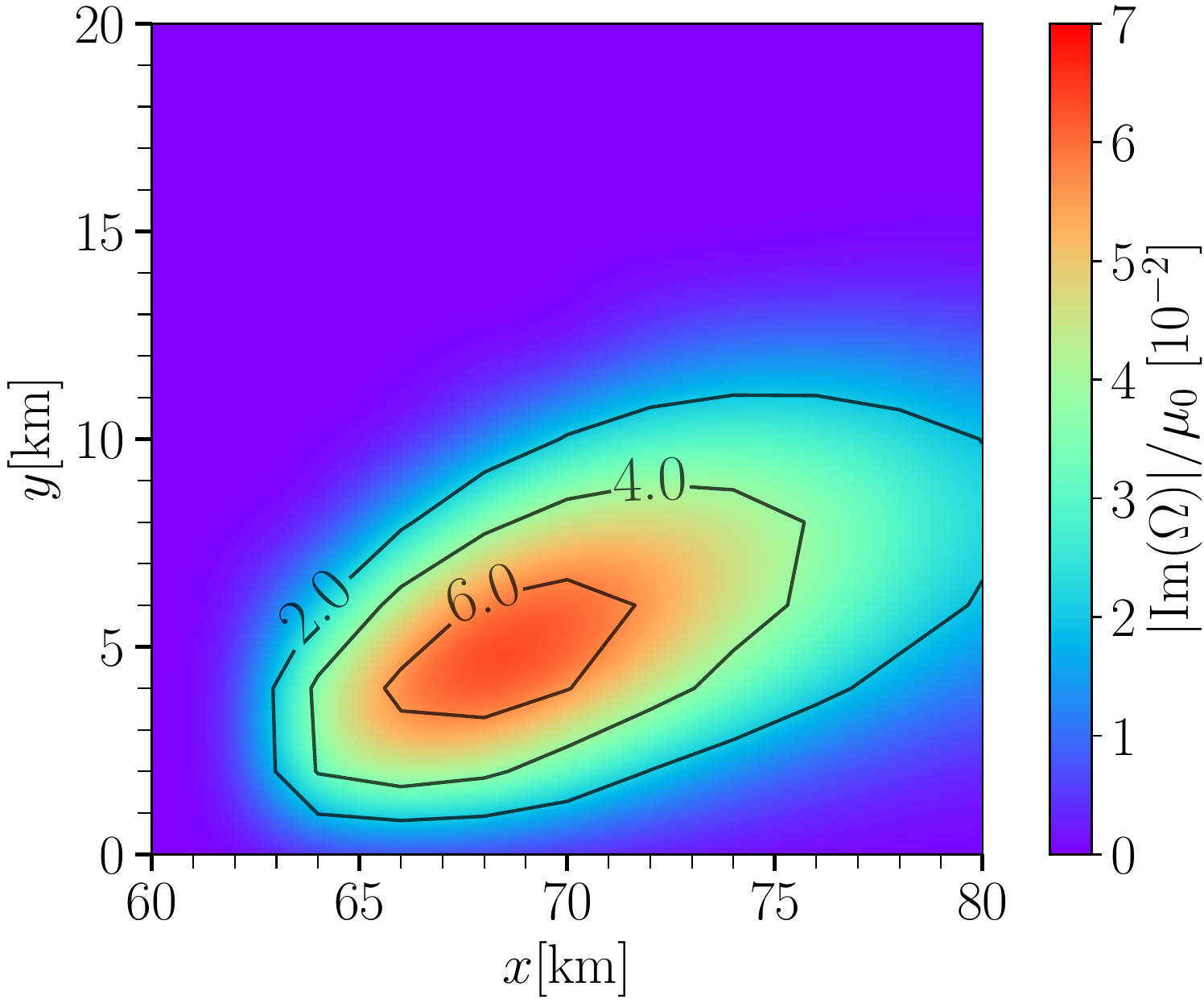}
\caption{{\it Left:} Contour plot of the instability parameter $\zeta$ (see Eq.~\ref{zeta:def}) in the 2D box in the absence of flavor conversions for the $\nu_e$ and $\bar\nu_e$ distributions in Fig.~\ref{fig:1}. The ELN crossings are significant just above the edges of the emitting sources. 
 {\it Right:} Contour plot of $|\mathrm{Im}(\omega)|/\mu_0$ for the homogeneous mode for the benchmark NS-disk model in one of the regions where the instability parameter is the largest (see dashed lines in the left panel). A maximum growth rate $|\mathrm{Im}(\omega)|= 0.07\mu_0 \simeq 2$~ns can be achieved. 
}
\label{fig:1a}
\end{figure}

The left panel of Fig.~\ref{fig:1a} shows a contour plot of the instability parameter in the absence of flavor conversions across our 2D box. One can see that $\zeta$ is large in the proximity of the edges of the neutrino emitting surfaces ($x \simeq 15, 65$~km and $y \in [0,15]$~km) and it gradually decreases as we move away from the sources, since the (anti)neutrino gas dilutes and the ELN crossings become less prominent. As a consequence, and by taking into account that $\mu(|\vec{x}|)$ decreases as we move away from $\mathcal{S}_{\nu}$ and $\mathcal{S}_{\bar\nu}$ (see Eq.~\ref{eq:mudef}), we should expect fast pairwise conversions to possibly occur where the $\zeta$ parameter is larger. Also, it is worth noticing that $\zeta$ is approximately zero in the central region of the emitting sources ($x \in [20,60]$~km and $y \in [0,15]$~km), this is mostly a consequence of the fact that we assume the neutrinospheres of $\nu_e$ and $\bar\nu_e$ to be coincident with each other, despite differing in width. Similarly to Ref.~\cite{Wu:2017qpc}, we expect to find a suppression of the flavor instabilities in the proximity of the emitting surfaces around the polar region ($\zeta$ is very small in our case) and a growth of the instabilities at larger distances from the source ($\zeta$ becomes larger).

Note that flavor conversions affect the (anti)neutrino angular distributions. Hence, the instability parameter shown in Fig.~\ref{fig:1a} can be dynamically modified by fast pairwise conversions. However, the plot provides with insights on the regions where flavor conversions may have larger effects, as we will see in Sec.~\ref{sec:nsm}.

\subsection{Linear stability analysis}\label{linearstability}
In order to explore the growth of the off-diagonal term in the density matrices, and therefore the development of fast pairwise conversions, we first rely on the linear stability analysis to analytically predict the growth rate of the flavor instabilities~\cite{Banerjee:2011fj,Raffelt:2013rqa}. Note that, given that we intend to focus on fast pairwise conversions, we assume $\omega=0$ in this section; such assumption is justified since $\omega \neq 0$ would mainly affect the non-linear regime of fast pairwise conversions~\cite{Shalgar:2020xns}. 

We linearize the EoM and track the evolution of the off-diagonal term through the following ansatz 
\begin{eqnarray}\label{eq:ansatz}
 \rho_{ex}(\theta) = Q(\theta)e^{-i\Omega t} \ \mathrm{and}\ 
 \bar{\rho}_{ex}(\theta) = \bar{Q}(\theta)e^{-i\Omega t} ,
\end{eqnarray}
where $\Omega = \gamma + i \kappa$ represents the collective oscillation frequency for neutrinos and antineutrinos. If Im$(\Omega) \neq 0$, then the flavor instability grows exponentially with rate $|$Im$(\Omega)|$, leading to fast pairwise conversions~\cite{Izaguirre:2016gsx}. Note that we look for temporal instabilities for the homogeneous mode ($\vec{k}=0$), as these are the ones possibly leading to fast pairwise conversions in extended regions~\cite{Wu:2017qpc}; by adopting a similar disk setup, Ref.~\cite{Wu:2017qpc} found that spatial instabilities occur in much smaller spatial regions than the temporal instabilities.

The off-diagonal component of Eq.~\ref{eom1} is 
\begin{eqnarray}\label{eom_lin}
 i \frac{\partial}{\partial t}\rho_{ex}(\theta) &=& H_{ee}(\theta)\rho_{ex}(\theta) + H_{ex}(\theta)\rho_{xx}(\theta) - \rho_{ee}(\theta)H_{ex}(\theta) - \rho_{ex}(\theta)H_{xx}(\theta) \nonumber \\ 
 &=& H_{ee}(\theta)\rho_{ex}(\theta) - \rho_{ee}(\theta)H_{ex}(\theta)\ ,
\end{eqnarray}
where we have assumed $\rho_{xx} (t=0~\mathrm{s}) = \bar\rho_{xx}(t=0~\mathrm{s}) = 0$. By substituting Eq.~\ref{eq:ansatz} in the equation above and solving for $Q(\theta)$, we obtain 
\begin{eqnarray}\label{eq:Qtheta}
 Q(\theta) = \frac{ \rho_{ee}(\theta) \int d\theta^{\prime} [ Q(\theta^{\prime})-\bar{Q}(\theta^{\prime}) ] [1-\cos{(\theta-\theta^{\prime})}] } { -\frac{\Omega}{\mu} + \int d\theta^{\prime} [ \rho_{ee}(\theta^{\prime})-\bar{\rho}_{ee}(\theta^{\prime}) ] [1-\cos{(\theta-\theta^{\prime})}] }\ .
\end{eqnarray}
A similar procedure follows for $\bar{Q}_{\theta}$ (see Eqs.~\ref{eom2} and \ref{eq:ansatz}). Then, combining the expressions for $Q(\theta)$ and $\bar{Q}(\theta)$, we have 
\begin{eqnarray}\label{eq:QmQbar1}
 Q(\theta)-\bar{Q}(\theta) = \int d\theta^{\prime} \Big(\frac{\rho_{ee}(\theta)-\bar{\rho}_{ee}(\theta)}{-\frac{\Omega}{\mu} + A(\theta)}\Big) [ Q(\theta^{\prime})-\bar{Q}(\theta^{\prime}) ] [1-\cos{(\theta-\theta^{\prime})}]\ ,
\end{eqnarray}
where $A(\theta) = \int d\theta^{\prime} [ \rho_{ee}(\theta^{\prime})-\bar{\rho}_{ee}(\theta^{\prime}) ] [1-\cos{(\theta-\theta^{\prime})}]$. 
From the equation above, it must be true that 
\begin{eqnarray}\label{eq:QmQbar2}
 Q(\theta)-\bar{Q}(\theta) = \left[\frac{\rho_{ee}(\theta)-\bar{\rho}_{ee}(\theta)}{-\frac{\Omega}{\mu} + A(\theta)}\right] (a - b\cos{\theta} - c\sin{\theta})\ ,
\end{eqnarray}
where $a,b,c$ are unknown coefficients. By substiting Eq.~\ref{eq:QmQbar2} in Eq.~\ref{eq:QmQbar1}, we obtain a system of equations for the coefficients $a$, $b$, and $c$. Since the variable $\theta^{\prime}$ is a dummy variable, we replace it by $\theta$: 
\begin{eqnarray}\label{eq:system}
\begin{bmatrix}
a \\
b \\
c
\end{bmatrix}
=
\begin{bmatrix}
\mathcal{I}[1] & - \mathcal{I}[\cos{\theta}] & - \mathcal{I}[\sin{\theta}] \\
\mathcal{I}[\cos{\theta}] & - \mathcal{I}[\cos{\theta}^2] & - \mathcal{I}[\cos{\theta}\sin{\theta}] \\
\mathcal{I}[\sin{\theta}] & - \mathcal{I}[\cos{\theta}\sin{\theta}] & - \mathcal{I}[\sin{\theta}^2]
\end{bmatrix}
\begin{bmatrix}
a \\
b \\
c
\end{bmatrix}
=
\mathrm{M}
\begin{bmatrix}
a \\
b \\
c
\end{bmatrix} \ ,
\end{eqnarray}
where the functional $\mathcal{I}[f]$ is
\begin{eqnarray}
 \mathcal{I}[f] = \int d\theta \left[\frac{\rho_{ee}(\theta)-\bar{\rho}_{ee}(\theta)}{-\frac{\Omega}{\mu} + A(\theta)}\right] f(\theta)\ .
\end{eqnarray}
The system of equations
has a not trivial solution if
\begin{eqnarray}\label{eqn:det}
 \mathrm{det}( \mathrm{M} - 1 ) = 0\ . 
\end{eqnarray}
The latter equation is polynomial in the frequency $\Omega$. To search for instabilities, we need to look for the solutions with $\mathrm{Im}(\Omega) = \kappa \neq 0$. We then use the SciPy module~\cite{SciPy} in Python to find the roots numerically.

The right panel of Fig.~\ref{fig:1a} shows the growth rate, $|\mathrm{Im}(\Omega)|/\mu_0$, for a region of our 2D box where the instability parameter is the largest (see the highlighted region in the left panel of Fig.~\ref{fig:1a}). In the region of the 2D box corresponding to the edges of $\mathcal{S}_{\nu}$, $|\mathrm{Im}(\Omega)|/\mu_0 \simeq 0.01$--$0.06$; if we compare our findings to the ones reported in the top panel of Fig.~3 of Ref.~\cite{Wu:2017qpc}, we obtain a roughly comparable growth rate of the flavor instability. We should highlight that we assume the $\nu_e$ and $\bar\nu_e$ neutrinospheres to be exactly coincident with each other (although having different widths) while a two-disk model was considered in Ref.~\cite{Wu:2017qpc}; this quantitatively affects the depth of the ELN crossings in the polar region above the remnant in the proximity of the source. 
We also note that we model differently the edges of the (anti)neutrino sources and the (anti)neutrino angular distributions with respect to Ref.~\cite{Wu:2017qpc} and this causes differences in the shape of the unstable regions above the NS-disk remnant.

\section{Flavor evolution above the NS-disk remnant}
\label{sec:nsm}
Most of the existing work in the context of neutrino flavor conversions above the remnant disk focuses on exploring the phenomenology of slow collective oscillations and the matter-neutrino resonance~\cite{Malkus:2012ts,Malkus:2014iqa,Wu:2015fga,Zhu:2016mwa,Frensel:2016fge,Tian:2017xbr,Shalgar:2017pzd}. The only existing literature on fast pairwise conversions in merger remnants relies on the linear stability analysis to explore whether favorable conditions for fast conversions exist above the remnant disk~\cite{Wu:2017qpc,Wu:2017drk}, as also discussed in Sec.~\ref{linearstability}. In this section, we present the results of the numerical evolution in the non-linear regime of fast pairwise conversions above the NS-disk remnant and discuss the implications for the merger physics. We then generalize our findings by exploring the parameter space of the possible $\nu_e$--$\bar\nu_e$ asymmetries expected above the NS-disk remnant and the relative ratio between the size of the $\nu_e$ and $\bar\nu_e$ sources.

\subsection{Numerical implementation}
\label{sec:numimpl}
We solve the EoM introduced in Sec.~\ref{sec:fast} for the box setup described in Sec.~\ref{sec:setup} by following the procedure outlined in Sec.~3.2 of Ref.~\cite{Shalgar:2019qwg}. In the numerical runs, we adopt $N_x = N_y = 50$ number of bins for the $x-y$ grid and $N_\theta = 300$ angular bins to ensure numerical convergence.

\begin{figure}
\centering
\includegraphics[width=0.7\textwidth]{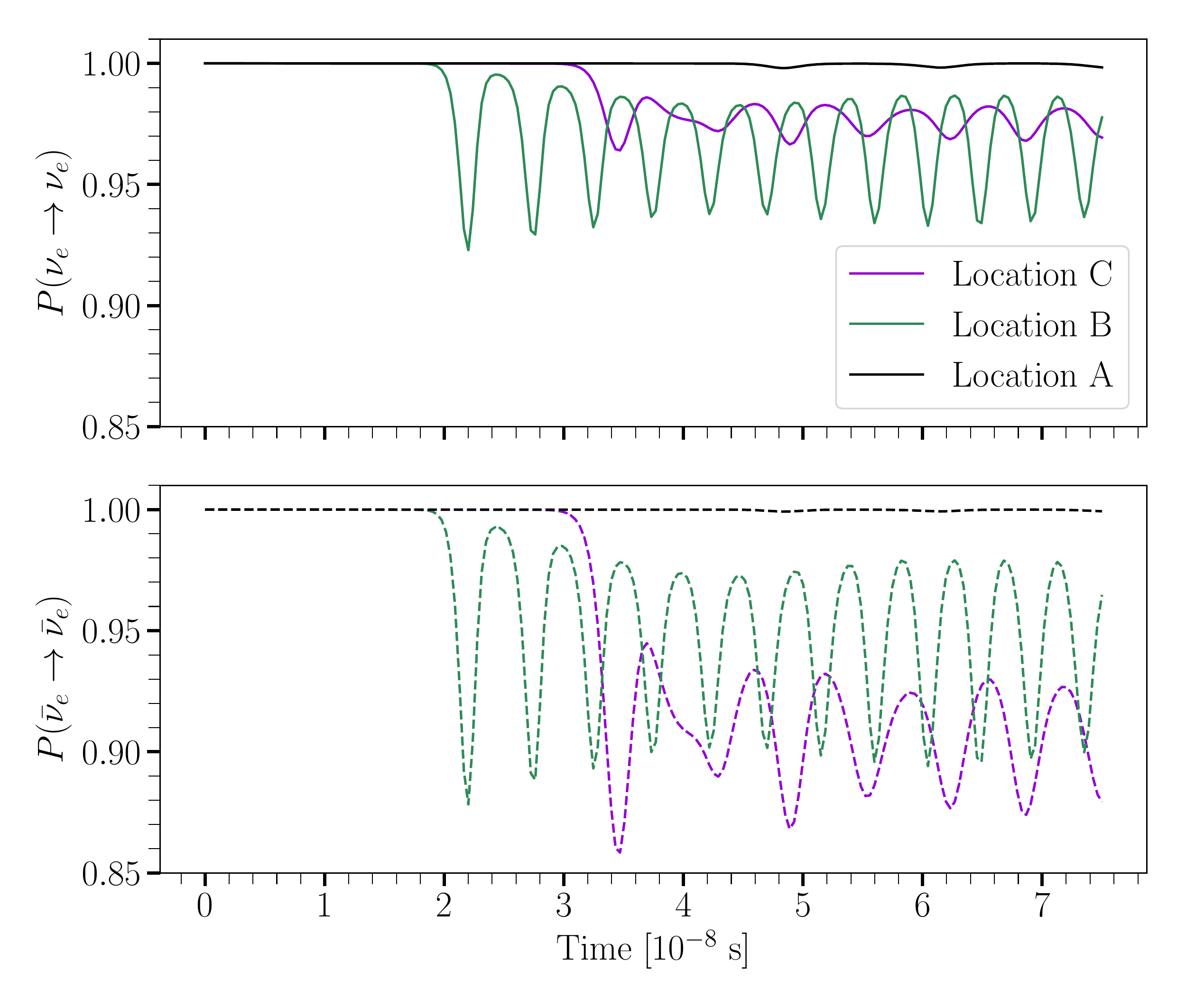}
\caption{Temporal evolution of the survival probabilities of $\nu_e$ (top) and $\bar{\nu}_e$ (bottom) for the three selected locations A, B and C shown in Fig.~\ref{fig:1} (see Eq.~\ref{eq:Pee}). After a certain time, $\Delta t \simeq \mathcal{O}(10^{-7})$~s, the survival probabilities have reached a steady-state configuration. Minimal flavor mixing is achieved for all three locations.
}
\label{fig:9}
\end{figure}

In order to quantify the amount of flavor mixing, we introduce the angle integrated survival probabilities
\begin{eqnarray}\label{eq:Pee}
P(\nu_e \rightarrow \nu_e) = \frac{ \int d\theta [\rho_{ee}(\vec{x},t) - \rho_{xx}(\vec{x},t=0 \ \mathrm{s})] }{ \int d\theta [\rho_{ee}(\vec{x},t=0 \ \mathrm{s}) - \rho_{xx}(\vec{x},t=0 \ \mathrm{s})] } \ ,
\end{eqnarray}
and similarly for $P(\bar{\nu}_e \rightarrow \bar{\nu}_e)$ with the replacement $\rho \rightarrow \bar{\rho}$. Figure~\ref{fig:9} shows the survival probabilities of $\nu_e$ and $\bar\nu_e$ as  functions of time for the three selected $(x,y)$ locations (A, B, and C) in the 2D box, see Fig.~\ref{fig:1}. One can easily see that fast pairwise conversions take some time to develop, but then they reach a ``steady-state'' configuration and the survival probability stabilizes, despite smaller scale oscillations, without changing dramatically. 

In the presence of flavor conversions, for each $(x,y)$ point in the 2D box, flavor conversions develop on a time scale shorter than the advective time scale~\cite{Shalgar:2019qwg}. To take into account the (anti)neutrino drifting through the 2D box,  for each $(x,y)$ location in the 2D box, we translate the time-averaged neutrino and antineutrino density matrices from each spatial bin to the neighboring bins after a time $\Delta t \simeq \mathcal{O}(10^{-7})$~s, i.e. after the flavor conversion probability in $(x,y)$ has reached a steady-state configuration; we keep all the parameters within each spatial bin unchanged, except for following the flavor conversions for smaller time intervals. 
This procedure is implemented in an automated fashion as described in Appendix~\ref{appendix:algorithm} and it naturally allows to recover the flavor configuration shown in Fig.~\ref{fig:1} in the absence of flavor conversions. We stress that our procedure automatically allows to take into account the dynamical evolution of the angular distributions as a function of time, due to neutrinos streaming from the neighboring bins.

As seen in Fig.~\ref{fig:9} the (anti)neutrino occupation numbers oscillate around an average value after the system has reached the non-linear regime. In an astrophysical system, at a given point in space, only the time-averaged occupation numbers are the relevant quantities as  long as the size of the region over which neutrinos and antineutrinos are emitted is larger than the length scale over which neutrinos and antineutrinos oscillate. The aforementioned condition should always be satisfied above the remnant disk because of  the short time-scales over which fast flavor conversions occur.

It is worth noticing that, while Fig.~\ref{fig:1} represents the resultant angular distributions of $\nu_e$ and $\bar\nu_e$ in the absence of flavor conversions across the 2D box, by streaming the oscillated (anti)neutrinos to their neighboring bins, we also modify the angular distributions dynamically. In Ref.~\cite{Shalgar:2019qwg}, it was shown the neutrino advection smears the ELN crossings hindering the development of fast pairwise conversions; such an effect would eventually become efficient on time scales longer than $\Delta t$, i.e.~after the steady-state configuration has been reached in our 2D box. Moreover, the ELN crossings in our system are assumed to be self-sustained in time because of the disk geometry and its protonization.

For a selection of points close to $\mathcal{S}_{\nu}$ and $\mathcal{S}_{\bar{\nu}}$, we have also tested that the growth rate obtained from the linear stability analysis in Sec.~\ref{linearstability} perfectly matches the linear regime of the numerical solution. In particular, for the point B in Fig.~\ref{fig:1}, we find $|$Im$(\Omega)|/\mu = 0.0489$ by numerically solving the EoM and $|$Im$(\Omega)|/\mu = 0.049$ by relying on the stability analysis. As discussed in Ref.~\cite{Shalgar:2020xns}, the dependence of the linear growth rate on $\omega$ is overall negligible.

\subsection{Results}
\label{sec:results}

Figure~\ref{fig:9} shows the temporal evolution of the $\nu_e$ and $\bar\nu_e$ survival probabilities as discussed in Sec.~\ref{sec:numimpl}. Even though flavor unstable solutions are predicted to exist almost at any location above the disk of the remnant and the linear stability analysis suggests a large growth rate, as shown in Fig.~\ref{fig:1a}, our results show that fast pairwise conversions lead to a few percent variation in the flavor transition probability. 
At most, an average value of $P(\nu_e \rightarrow \nu_e)\simeq 0.95$ is obtained in our runs; while the survival probability for antineutrinos can reach slightly lower values, $P(\bar\nu_e \rightarrow \bar\nu_e) \simeq 0.90$, due to the lepton number conservation.

\begin{figure}
\centering
\includegraphics[width=0.7\textwidth]{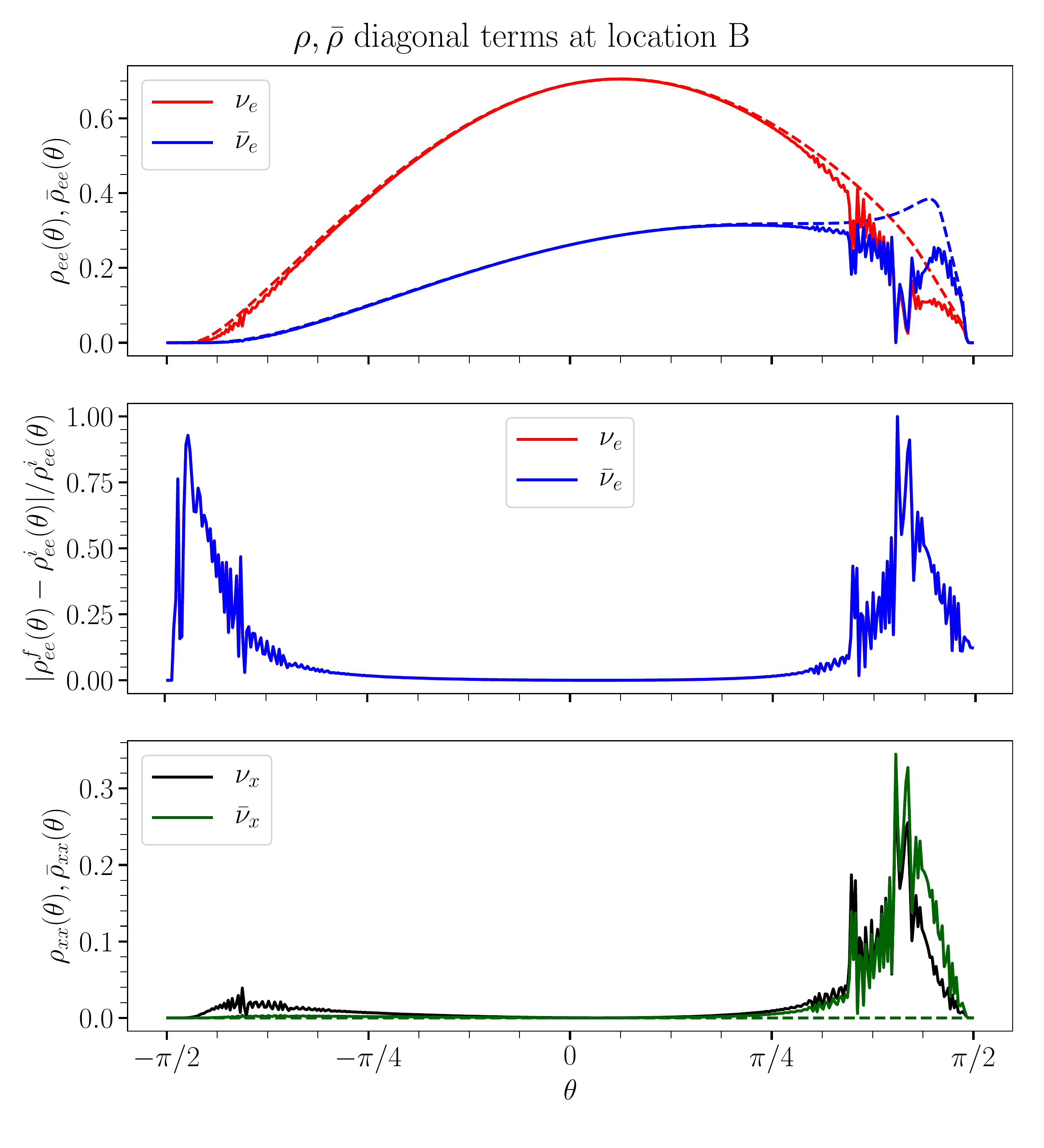}
\caption{{\it Top:} Density matrix elements $\rho_{ee}(\theta)$ (in red) and $\bar{\rho}_{ee}(\theta)$ (in blue) as functions of the emission angle $\theta$ at $t=0$~s (dashed curves, $i$) and at $t=7.5 \times 10^{-8}$~s (solid curves, $f$) at the selected location B of Fig.~\ref{fig:1}. The density matrix elements are normalized to $\rho_{\mathrm{tot}}$.
\textit{Middle:} Relative difference between the final state $(f)$ and initial state $(i)$ of $\rho_{ee}(\theta)$ (red) and $\bar{\rho}_{ee}(\theta)$ (blue), respectively.
\textit{Bottom:} Same as the top panel, but for the density matrix elements $\rho_{xx}(\theta)$ (black) and $\bar{\rho}_{xx}(\theta)$ (green).
Fast pairwise conversions only affect the tail of the angular distributions of $\nu_e$ and $\bar\nu_e$ inducing minimal changes. 
}
\label{fig:4}
\end{figure}

To better explore the development of fast pairwise conversions as a function of the emission angle, the top panel of Fig.~\ref{fig:4} shows $\rho_{ee}$ and $\bar\rho_{ee}$ (respectively proportional to the $\nu_e$ and $\bar\nu_e$ number densities) as functions of the emission angle $\theta$ for the observer located in B in Fig.~\ref{fig:1}. The angular distributions are displayed at $t=0$~s ($i$, dashed lines) and at $t=7.5 \times 10^{-8}$~s ($f$, solid lines) when the flavor conversions have reached a steady-state configuration (see Fig.~\ref{fig:9} and Appendix~\ref{appendix:algorithm}). Initially, the ELN crossings are large for $\theta \in [\pi/4,\pi/2]$; this triggers fast pairwise conversions of $\nu_e$ and $\bar{\nu}_e$. 
As $t$ increases, the angular distributions of $\nu_e$ and $\bar\nu_e$ most prominently evolve around  the angular bins in the proximity of the ELN crossing, as highlighted in the middle panel of Fig.~\ref{fig:4}, until the density matrices reach a stationary value. 
In the bottom panel of Fig.~\ref{fig:4}, one can see that newly formed $\nu_x$ and $\bar\nu_x$ angular distributions peak in a very narrow $\theta$ interval where the ELN crossings occur. As a result, $\nu_x$ ($\bar\nu_x$) will predominantly propagate outwards and away from the remnant symmetry axis, thus having a marginal impact on the polar region of the system.

\begin{figure}
\centering
\includegraphics[width=0.82\textwidth]{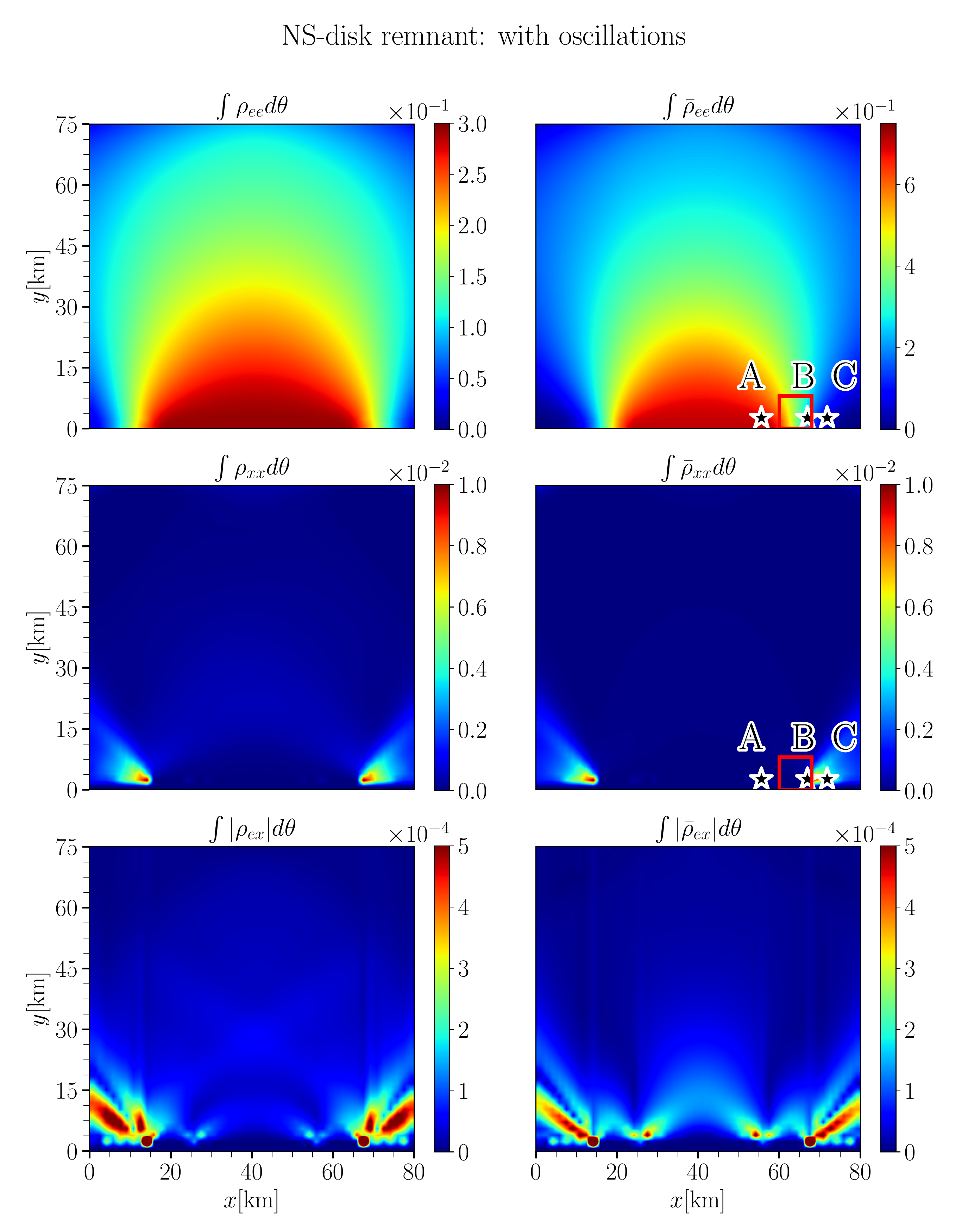}
\caption{Contour plots of the angle-integrated elements of the density matrices $\rho(\vec{x},\theta,t)$ (one the left) and $\bar{\rho}(\vec{x},\theta,t)$ (on the right) for the NS-disk remnant configuration normalized to the maximum total particle number within the box i.e. $\int d\theta (\rho_{ee}+\bar{\rho}_{ee}+2\rho_{xx})$, after $\Delta t =10^{-7}$~s, after the flavor distribution has reached a steady-state configuration. Three selected locations (A, B, and C) are highlighted (see Fig.~\ref{fig:1}). The red box defines the region scanned with higher spatial resolution, see Appendix~\ref{appendix:spatial} for details. As also shown in Fig.~\ref{fig:1a}, fast pairwise conversions are more prominent near the edges of the neutrino sources. A small amount of $\nu_x$'s and $\bar\nu_x$'s is produced through fast pairwise conversions within the narrow opening angles at the edges of the neutrino surfaces. The ELN crossings are almost vanishing along the axis of symmetry leading to practically no flavor conversions in the polar region above the NS-disk remnant. }
\label{fig:16}
\end{figure}

Figure~\ref{fig:16} summarizes our findings across the 2D box by displaying contours of the angle-integrated density matrix elements for neutrinos (on the left) and antineutrinos (on the
\newpage
\noindent
 right) for the NS-disk remnant configuration when the steady-state configuration for flavor conversions is reached. 
The top panels are almost identical to the ones in Fig.~\ref{fig:1} because of the overall small amount of flavor conversions despite the large instability parameter and growth rate (see Fig.~\ref{fig:1a}). From the middle and the bottom panels, we can clearly see that  flavor conversions occur in the region at the edges of the emitting surfaces where $\zeta$ is larger, but they have a negligible role in the polar region above the remnant disk where the neutrino-driven wind nucleosynthesis could be affected~\cite{Wu:2017drk}.

For completeness,  Appendix~\ref{appendix:spatial} includes results of a high resolution run performed in the red box in Fig.~\ref{fig:16}. The overall amount of flavor conversions is comparable in the  low and high resolution simulations; for this reason, we have chosen to rely on simulations with lower resolution in order to explore a larger region above the remnant disk.

By comparing Figs.~\ref{fig:1a} and Fig.~\ref{fig:16}, we conclude that the high linear growth rate of fast 
  pairwise conversions does not imply an overall large flavor conversion in the non-linear regime. However, we should stress that ours is the first numerical study of fast pairwise conversions
above the merger remnant in the non-linear regime; as such, for the sake of simplicity, we have neglected the collision term in EoM. The collision term may potentially play a significant role, also because it generates a backward flux of (anti)neutrinos~\cite{Dasgupta:2016dbv,Abbar:2017pkh}, which is neglected in our setup. A better refined modeling of the neutrino conversion physics may affect the flavor outcome with implications for the $r$-process nucleosynthesis~\cite{Wu:2017qpc,Wu:2017drk}.

\subsection{Results for other NS-disk configurations}
\begin{figure}[b]
\centering
\includegraphics[width=0.7\textwidth]{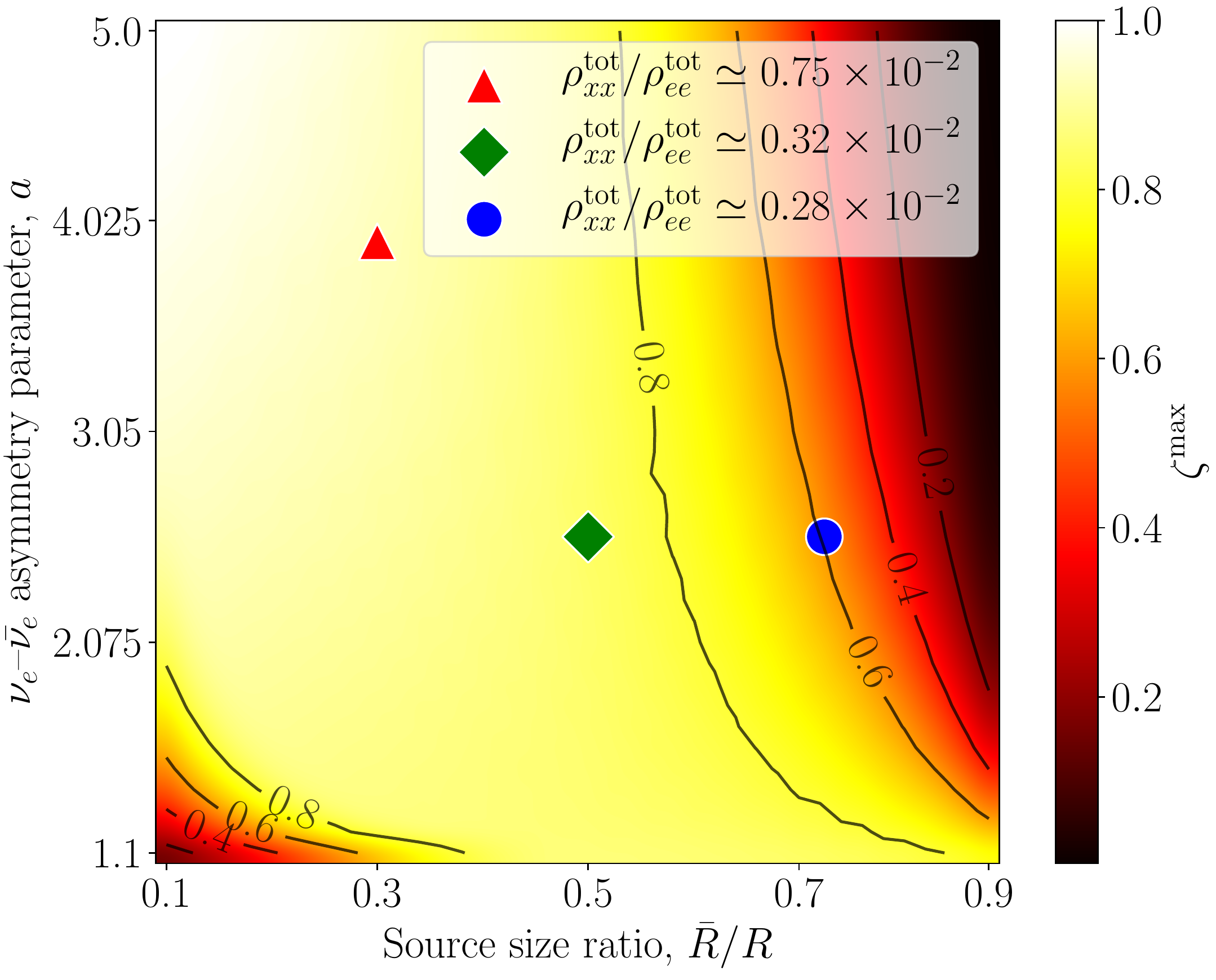}
\caption{Contour plot of the maximal value of the instability parameter (Eq.~\ref{zeta:def}) in the parameter space defined by the relative ratio between the $\bar\nu_e$ and $\nu_e$ source sizes ($\bar{R}/R$) and the $\nu_e$--$\bar\nu_e$ asymmetry parameter ($a$). For each point in the parameter space, corresponding to a NS-disk configuration, the maximum value of $\zeta$ is computed in the absence of oscillations. In order to gauge the overall amount of flavor conversion, the three colored diamonds represent three NS-disk configurations for which we have tracked the flavor evolution numerically. The transition probability is reported in the legend for each of the three selected configurations. A slightly larger transition probability is obtained for smaller $\bar{R}/R$ ratios. 
}
\label{fig:29}
\end{figure}
In order to gauge how our findings are modified for different configurations of the NS-disk remnant, in this section we vary the $\nu_e$--$\bar\nu_e$ asymmetry parameter, $a$ (see Eq.~\ref{case2}) within the range allowed from hydrodynamical simulations~\cite{Frensel:2016fge}, as well as the relative ratio between the sizes of $\mathcal{S}_{\bar\nu}$ and $\mathcal{S}_{\nu}$ ($\bar{R}/R$). 

Figure~\ref{fig:29} shows a contour plot of the maximum of the instability parameter $\zeta$ (Eq.~\ref{zeta:def}, computed in the absence of flavor conversions) computed across the 2D box for each ($\bar{R}/R, a)$. The markers in Fig.~\ref{fig:29} highlight three disk configurations that we have evolved numerically; the NS-disk configuration introduced in Sec.~\ref{sec:setup} is correspondent to $(\bar{R}/R, a) = (0.75, 2.4)$ (blue circle in Fig.~\ref{fig:29}). The conversion probability in the steady-state regime is $P(\nu_e \rightarrow P_{\nu_x}) \simeq 0.02$ for the NS-disk configuration introduced in Sec.~\ref{sec:setup} and tends to become larger for a smaller relative ratio of $\bar{R}/R$ [green diamond, $P(\nu_e \rightarrow P_{\nu_x}) \simeq 0.04$] and even more for the red triangle, 
where the instability parameter is maximal [$P(\nu_e \rightarrow P_{\nu_x}) \simeq 0.06$], which however probably corresponds to an extreme $\bar{R}/R$ ratio not realizable in astrophysical environments. 
Our findings suggest that flavor equilibration due to fast pairwise conversions is never achieved in our setup, despite the large growth rate predicted by the linear stability analysis. In addition, the regions in the 2D box that are most unstable are located in the proximity of the edges of the neutrino source, and no flavor conversions occur in the polar region above the NS-disk.

\section{Flavor evolution above the BH-disk remnant}
\label{sec:BHdisk}
We now extend our exploration of the phenomenology of fast pairwise conversions to the BH-torus configuration. In this case, the neutrino (antineutrino) source, $\mathcal{S}_{\nu}^{\mathrm{BH}}$ ($\mathcal{S}_{\bar{\nu}}^{\mathrm{BH}}$), is identical to the NS-disk remnant case except for an inner edge located at $R_{\mathrm{BH}} = 1/3 R$~\cite{Foucart:2015vpa} in correspondence of the innermost stable circular orbit, i.e.~the sources do not emit particles for $x \in [L/2 - R_{\mathrm{BH}}, L/2 + R_{\mathrm{BH}} ]$; all other model parameters are identical to the ones introduced in Sec.~\ref{sec:setup}. We observe that, in the case of the BH remnant, the neutrino and antineutrinos average energies are slightly higher than in the case of the massive NS remnant, see e.g.~\cite{Wu:2017drk,Just:2014fka}. However, since minimal variations in $\omega$ do not affect the final flavor configuration~\cite{Shalgar:2020xns}, we keep $\omega$ unchanged for simplicity.

\begin{figure}[h!]
\centering
\includegraphics[width=0.90\textwidth]{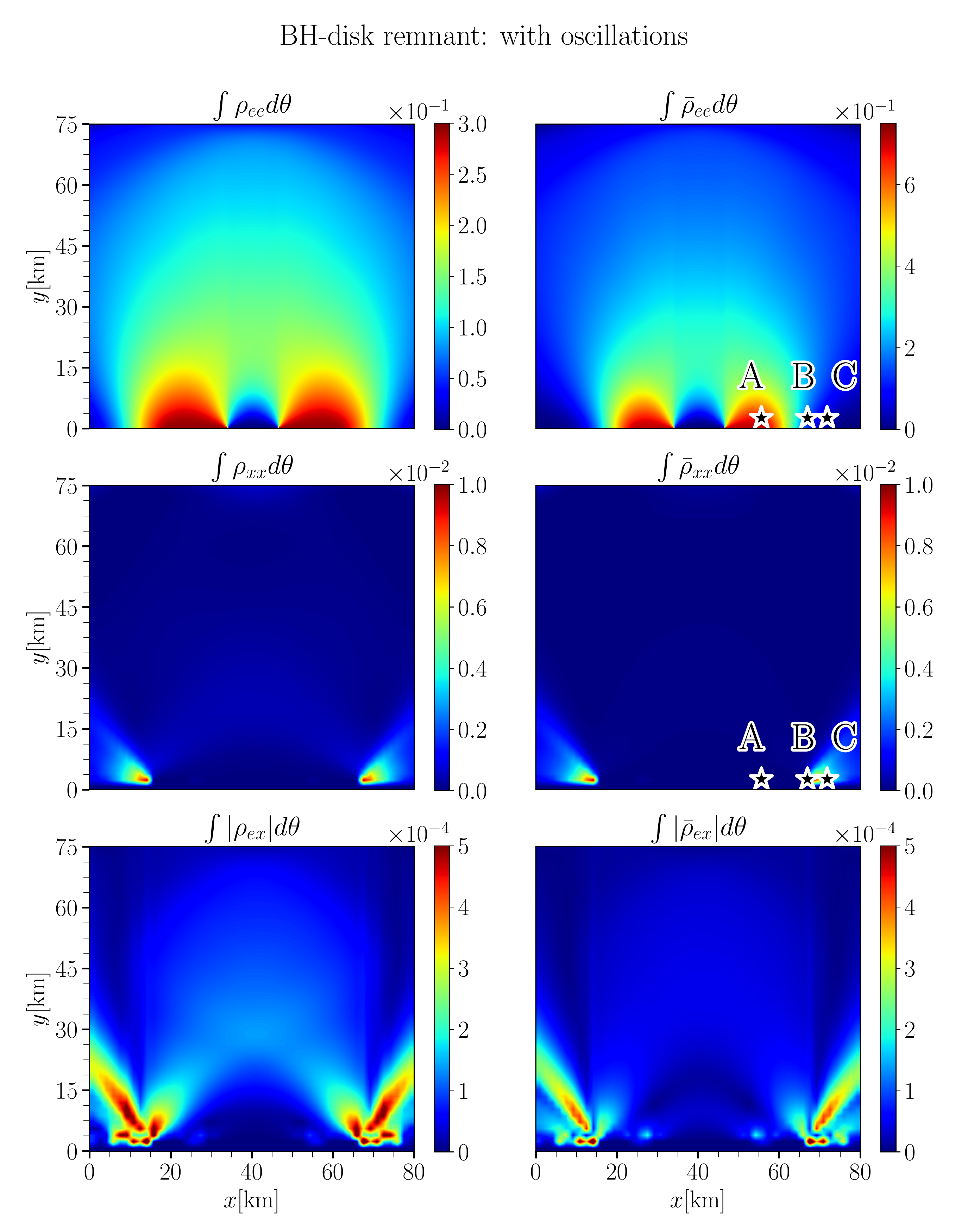}
\caption{Same as Fig.~\ref{fig:16} but for the BH-disk remnant configuration. Despite the differences in the source geometry, the final flavor configuration is comparable to the NS-disk remnant configuration. This suggests that the specific details of the source geometry do not lead to final state flavor configurations that are dramatically different. 
}
\label{fig:32b}
\end{figure}

Figure~\ref{fig:32b} shows the resultant angle-integrated density matrices for neutrinos and antineutrinos.
By comparing Figs.~\ref{fig:16} and \ref{fig:32b}, we can see that differences appear in the proximity of the inner source edges and just above the polar region, but the flavor distributions are comparable at larger distances from the source. Also, in this case, the most unstable regions appear in the proximity of the source external edges and minimal flavor conversions take place in the polar region, although more pronounced than for the NS-disk remnant configuration (see the bottom panels of Fig.~\ref{fig:32b} and Fig.~\ref{fig:16}). 

Our findings suggest that the flavor equipartition assumption adopted in Ref.~\cite{Wu:2017drk} to explore the implications on the nucleosynthesis of the heavy elements is difficult to achieve for our BH-disk configuration, despite the large growth rate predicted by the linear stability analysis (see Sec.~\ref{linearstability}). Similarly to the NS-disk configuration, the most unstable regions are located in the proximity of the edges of the neutrino sources.  However, in this case, minimal conversions occur in the polar region above the BH-disk where the neutrino wind may dominate the $r$-process outcome; these findings are in rough agreement with the unstable regions reported in the top panel of Fig.~7 of Ref.~\cite{Wu:2017drk} where a growth rate of the same order of the one plotted in the right panel of Fig.~\ref{fig:1a} was obtained.

It is worth noticing that Ref.~\cite{Wu:2017drk}, by studying the evolution of the BH-torus as a function of time, reported an excess of $\nu_e$ with respect to $\bar\nu_e$ in the polar region at late times (e.g., after $20$~ms) as a result of the dynamical evolution of the merger remnant. This effect is not taken into account in our simplified BH-disk, since we focus on a smaller time interval [$\Delta t \simeq \mathcal{O}(10^{-7})$~s] and we do not take into account modifications of the neutrino emission properties due to the dynamical evolution of the remnant.

\section{Outlook and conclusions}
\label{sec:conc}
Neutron star  merger remnants are dense in neutrinos, and the occurrence of electron lepton number (ELN) crossings between the angular distributions of $\nu_e$ and $\bar\nu_e$ seem to be ubiquitous as a natural consequence of the disk protonization and the source geometry. If fast pairwise conversions of neutrinos should occur, leading to flavor equipartition, this has been shown to lead to major consequences for the synthesis of the elements heavier than iron and the related kilonova observations~\cite{Wu:2017qpc,Wu:2017drk}. However, the existing literature on the subject focuses on predicting the existence of eventual flavor unstable regions by relying on the linear stability analysis.

In the light of the possible major implications for the source physics, for the first time, we solve the flavor evolution above the disk remnant in a (2+1+1) dimensional setup: two spatial coordinates, one angular variable, and time. This is the first computation of fast pairwise conversions above the merger disk in the non-linear regime. 

 For simplicity, we adopt a two-dimensional model with two coincident $\nu_e$ and $\bar\nu_e$ neutrinospheres, and a different size for the two sources.
 We look for the final steady-state configuration in the presence of fast pairwise conversions by neglecting the collisional term in the equations of motion and by mimicking a configuration where a massive neutron star sits at the center of the remnant disk (NS-disk configuration) and a configuration with a black hole remnant (BH-disk configuration). In addition, we scan the parameter space of the possible disk model parameters predicted by hydrodynamical simulations to test the robustness of our findings. 

We find that the most unstable regions favoring the occurrence of fast pairwise conversions are located in the proximity of the edges of the neutrino emitting surfaces. Only a minimal flavor change occurs in the polar region above the merger remnant in the BH-disk configuration, but flavor conversions are almost absent in the surroundings of the polar region in the NS-disk configuration. Fast pairwise flavor conversions are triggered early on and a steady-state configuration for the flavor ratio (modulo small high frequency modulations) is reached within $\mathcal{O}(10^{-7})$~s. 

Even though flavor unstable solutions are predicted to exist almost at any location above the disk of the remnant with a large growth rate, as already shown in the literature, our results point towards minimal flavor changes ($< 1\%$), which would suggest a negligible impact of fast pairwise conversions on the $r$-process nucleosynthesis. However, our findings should be taken with caution given the approximations intrinsic to our modeling. An interplay between fast and slow $\nu$--$\nu$ interactions in the context of the matter-neutrino resonance~\cite{Malkus:2012ts,Malkus:2014iqa,Wu:2015fga,Zhu:2016mwa,Frensel:2016fge,Tian:2017xbr,Shalgar:2017pzd} may occur, and a full solution of the flavor evolution in 3D may change  the flavor outcome yet again. 

This work  constitutes a major step forward in the exploration of fast pairwise conversions in the context of compact merger remnants from a quantitative perspective. Our findings suggest that a complete modeling of the neutrino flavor conversion physics should be taken into account in order to make robust predictions for the electromagnetic emission expected by the merger remnant and its aftermath.

\acknowledgments
We thank Meng-Ru Wu for useful comments on the manuscript. This project was supported by the Villum Foundation (Project No.~13164), the Danmarks Frie Forskningsfonds (Project No.~8049-00038B), the Knud H\o jgaard Foundation, and the Deutsche Forschungsgemeinschaft through Sonderforschungbereich
SFB~1258 ``Neutrinos and Dark Matter in Astro- and
Particle Physics'' (NDM).


\begin{appendix}

\section{The evolution algorithm}\label{appendix:algorithm}
In order to explore the flavor configuration achieved in our 2D box after a certain time $\Delta t$, we take into account neutrino advection in the EoM and aim to look for a ``steady-state'' flavor configuration, i.e.~for a configuration where the survival probability of (anti)neutrinos has reached a constant value as a function of time except for small oscillations around that value. In this Appendix, we describe the algorithm adopted to transport the (anti)neutrino gas through the advective operator ($\vec{v}\cdot\vec{\nabla}$) in the EoM. 

\begin{figure}[b]
\centering
\includegraphics[width=0.99\textwidth]{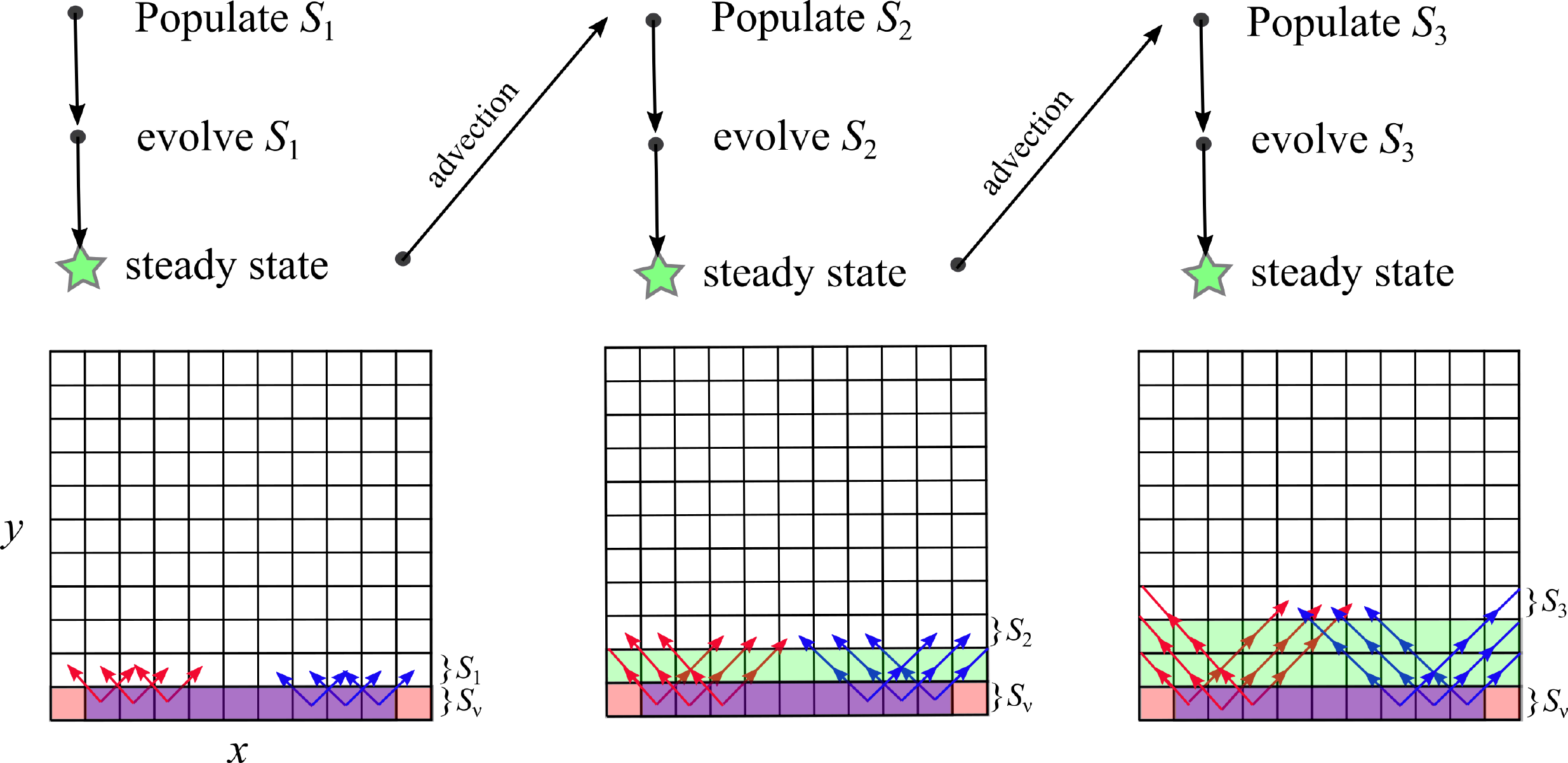}
\caption{Schematic representation of the algorithm implemented to determine the final steady-state configuration reached in our 2D system The neutrino and antineutrino decoupling regions, $\mathcal{S}_{\nu}$ and $\mathcal{S}_{\bar{\nu}}$, are plotted in red and blue, respectively. The regions $\mathcal{S}_{i}$ ($i=1,2,...,N_y)$ that reach a steady-state flavor configuration as a function of time are shown in green.
}
\label{fig:31}
\end{figure}

As sketched in Fig.~\ref{fig:31}, we evolve in time the flavor content in the different $\mathcal{S}_i$ regions in the box, individually and sequentially. We start from the one closest to the (anti)neutrino sources, $\mathcal{S}_{\nu}$ and $\mathcal{S}_{\bar{\nu}}$, through $\mathcal{S}_{N_y}$ at the opposite edge of the box. 

\begin{figure}
\centering
\includegraphics[width=0.70\textwidth]{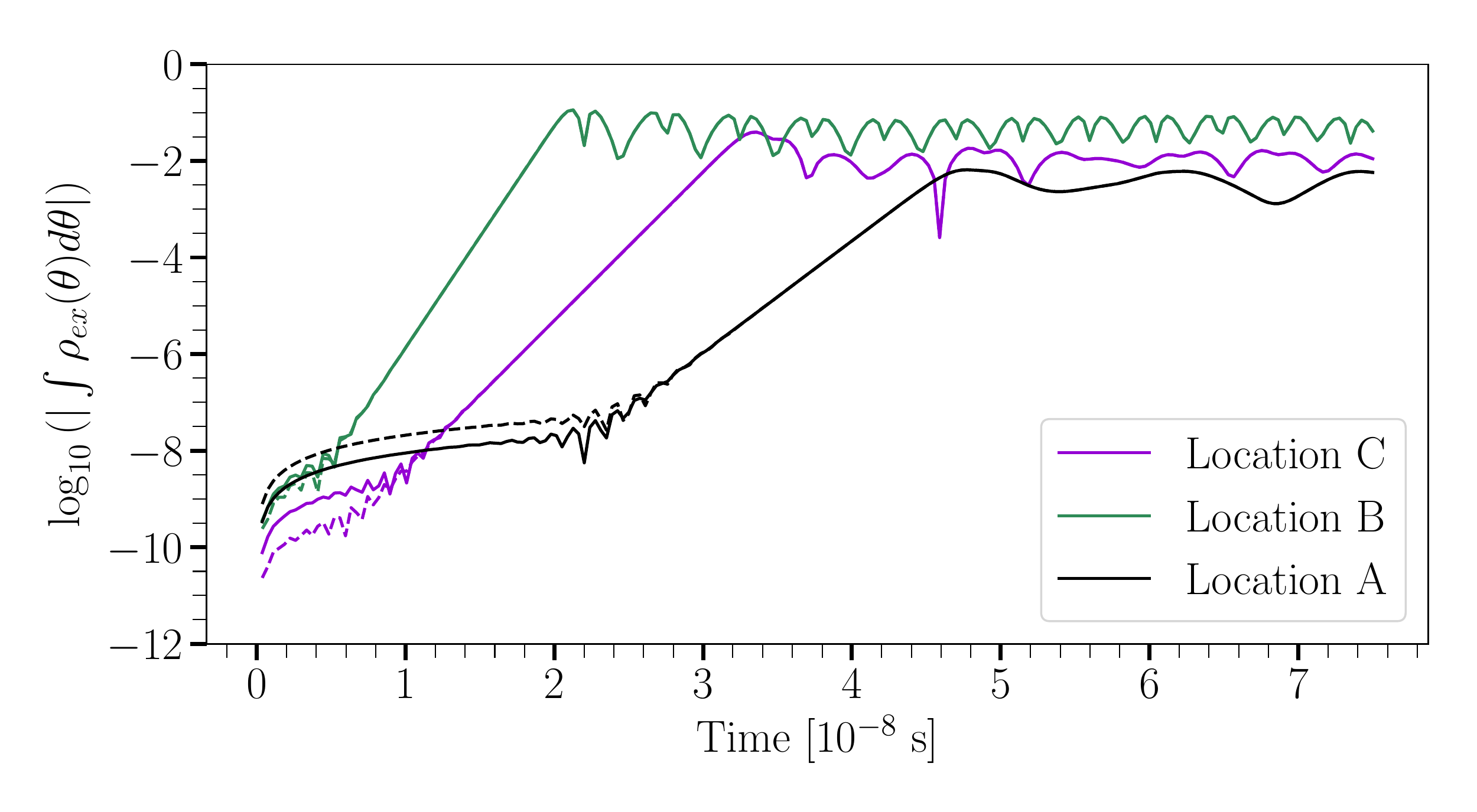}
\includegraphics[width=0.70\textwidth]{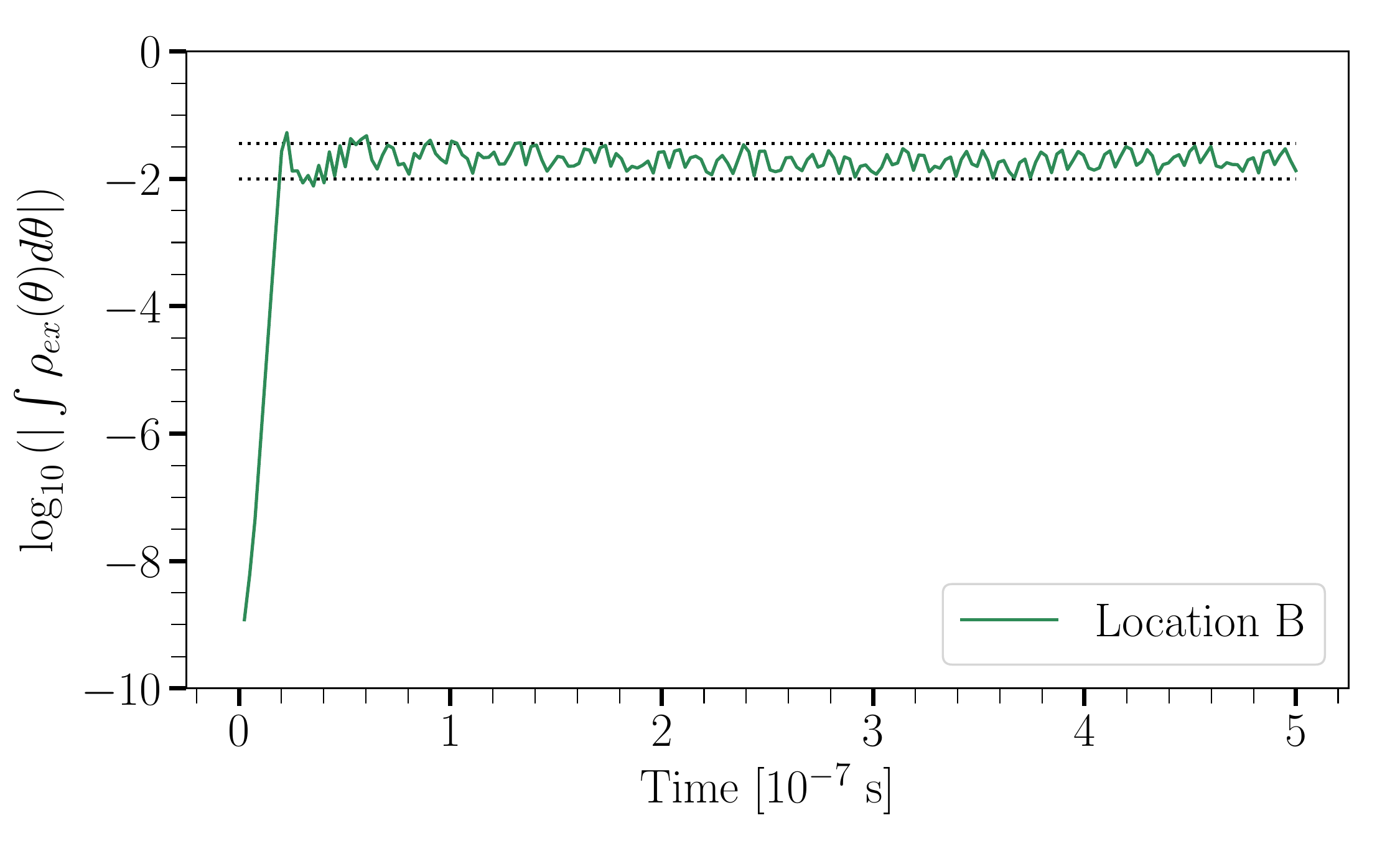}
\caption{\textit{Top:} Temporal evolution of $|\int \rho_{ex} d\theta|$ (solid) and $ |\int \bar{\rho}_{ex} d\theta|$ (dashed) matrix elements at the locations A, B and C, see Figs.~\ref{fig:1} and \ref{fig:9}. The exponential growth of the off-diagonal terms, and therefore, of the flavor instabilities, develops within a few ns. At a later stage, the system becomes highly non-linear and reaches an approximate steady-state.
\textit{Bottom}: Temporal evolution of $|\int \rho_{ex} d\theta|$ (location B), but tracking its temporal evolution for almost an order of magnitude longer. The dashed lines highlight the small variation of the transition probability, which allows to compute a steady-state flavor configuration.
}
\label{fig:7a}
\end{figure}

In our algorithm, the time-averaged density matrices are transported from $\mathcal{S}_i$ to $\mathcal{S}_{i+1}$, if a steady-state configuration of flavor conversions has been reached, e.g.~when the average values of $|\rho_{ex}|,|\bar{\rho}_{ex}|$ do not change more than a few percent ($\simeq 1\%$), as shown in Fig.~\ref{fig:7a}. This procedure is repeated in our algorithm to simulate the free streaming of neutrinos and antineutrinos through the box until the last region $S_{N_y}$ is reached. 

The sequential and pixel-by-pixel time evolution of the box is well motivated by physical arguments, namely, by the fact that $\nu$--$\nu$ interactions occur locally. A neutrino located at $(x,y)$ can only affect its nearest-neighbouring background neutrinos at $(x \pm \delta x, y \pm \delta y)$, where $\delta x,\delta y$ are infinitesimal displacements (the length of $\delta x, \delta y$ being set by the pixel length). In addition, the fact that we only stream neutrinos from the sources towards the opposite edges of the simulation box, and do not propagate them backwards, guarantees that a steady-state configuration is always achieved throughout the box.

\newpage

\section{Spatial resolution}\label{appendix:spatial}
 
In this appendix, we discuss the convergence of our results, especially regarding the spatial resolution adopted in this work.
In order to do this, we  solve the EoMs in a smaller box of $8 \times 8 \ \mathrm{km}^2$ (high resolution run), corresponding to the red box in Fig.~\ref{fig:16}, while maintaining the same number of grid points and all other input quantities unchanged. We follow the flavor evolution for $5 \times 10^{-7} \ \mathrm{s}$ by including neutrino advection at each time step. The red box has been located in one of the most unstable regions above the emitting surfaces, hence we expect that our test on the convergence of our results will provide an estimation of the largest possible error in the prediction of the flavor conversion probability.

\begin{figure}
 \centering
 \includegraphics[width=0.80\textwidth]{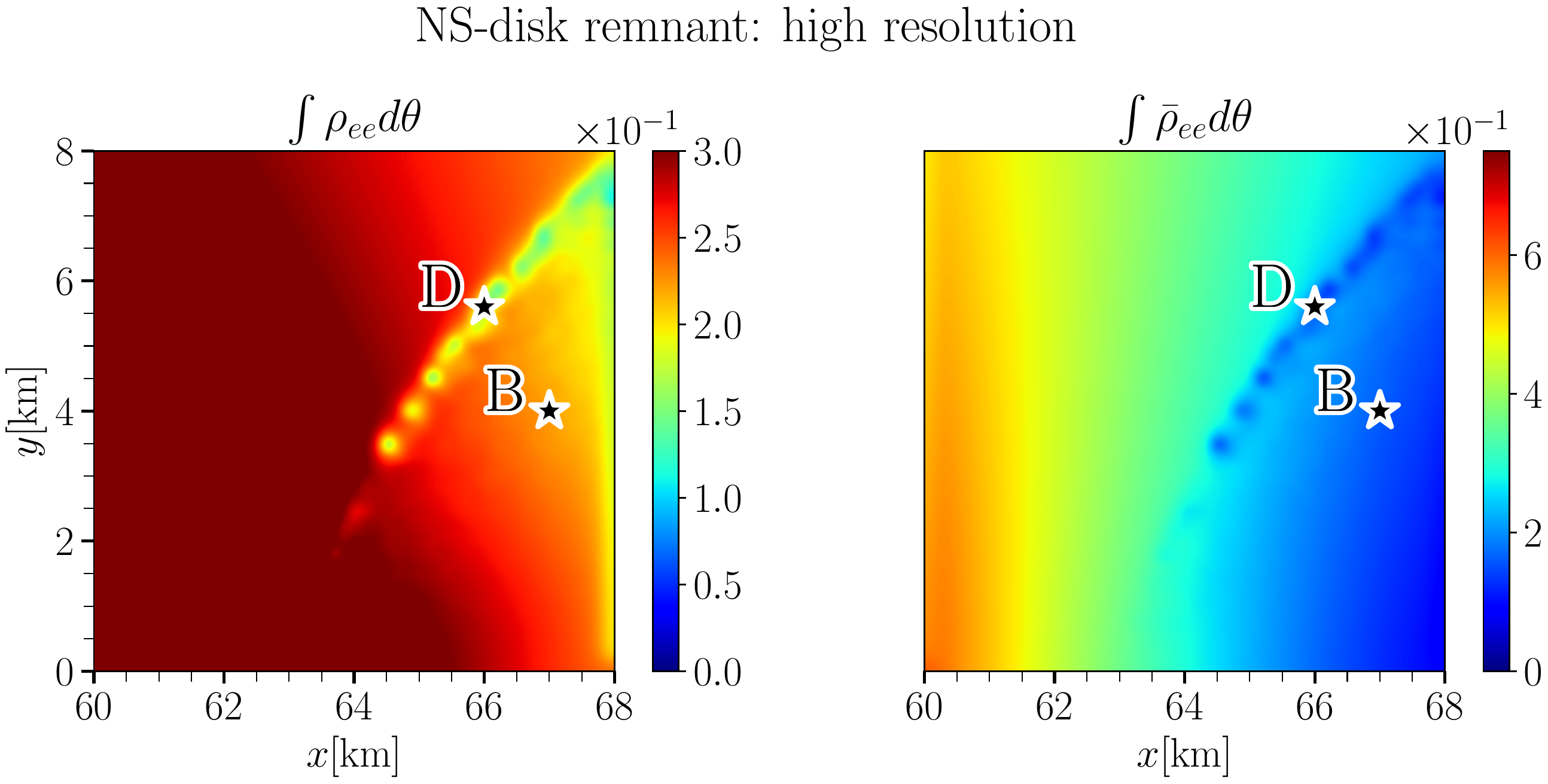}
  \includegraphics[width=0.80\textwidth]{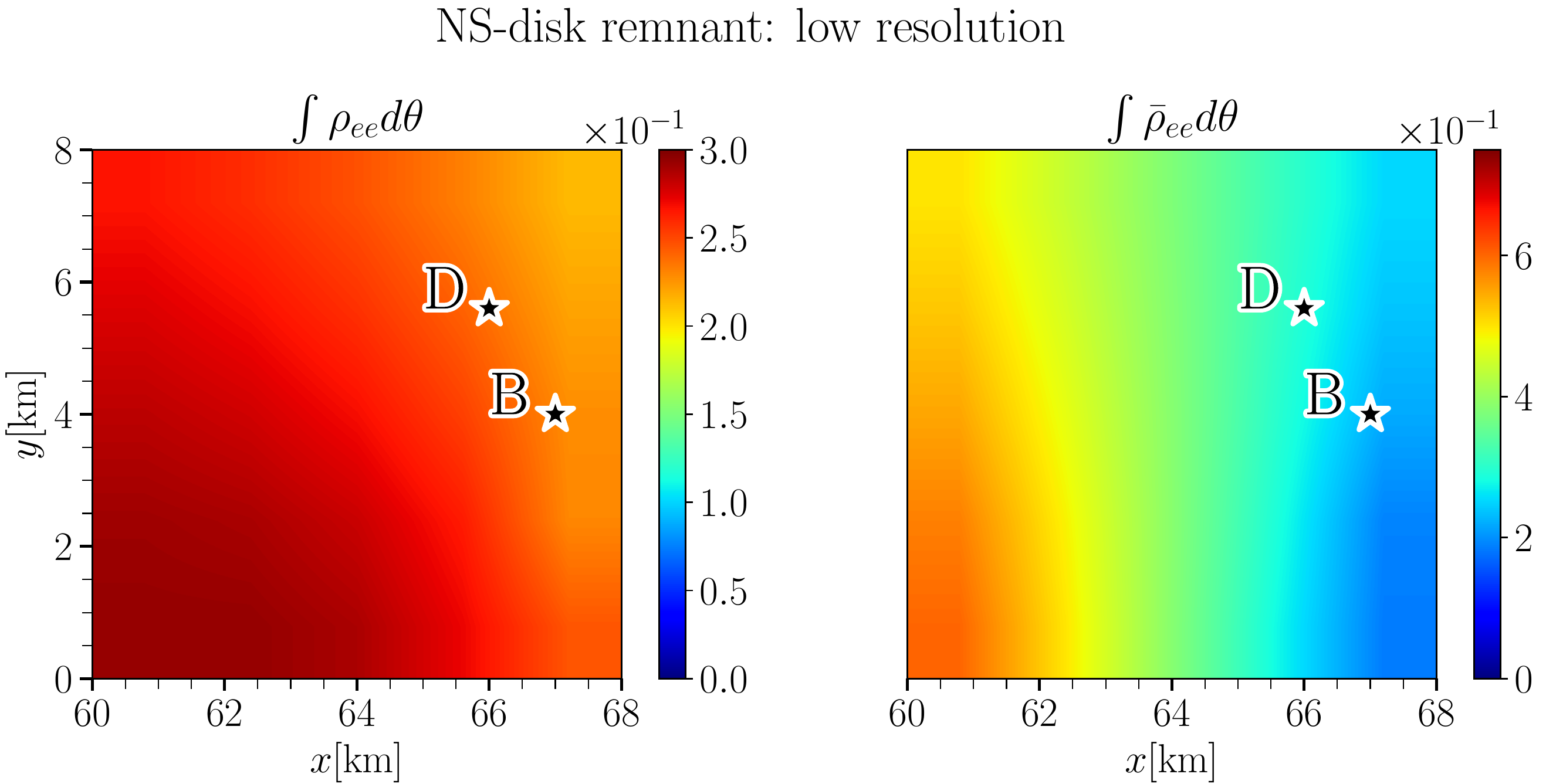}
 \caption{ 
 {Analogous to Fig.~\ref{fig:16}, contour plots of the angle-integrated elements of the density matrices, $\rho_{ee}(\vec{x}, \theta, t)$ (on the left) and $\bar\rho_{ee}(\vec{x}, \theta, t)$ (on the right) for the NS-disk configuration. The simulation domain is defined through a  $8 \times 8 \ \mathrm{km}^2$ spatial grid for the top panels (high resolution run) and a $80 \times 80 \ \mathrm{km}^2$ grid for the bottom panels (low resolution run, adopted throughout the paper); in both cases, the plotted region  corresponds to the red box in Fig.~\ref{fig:16}. Two selected locations (B and D) are used to inspect the temporal evolution of the survival probability, see Fig.~\ref{fig:14}.
The presence of small spatial structure does not affect the overall flavor evolution in the neighboring regions. }
}
 \label{fig:13}
 \end{figure}

Figure~\ref{fig:13} shows a contour plot of the angle integrated density matrix elements, $\rho_{ee}(\vec{x}, \theta, t)$  and $\bar\rho_{ee}(\vec{x}, \theta, t)$, at $5 \times 10^{-7} \ \mathrm{s}$ for the region highlighted by the red box in Fig.~\ref{fig:16}. The top panels have been obtained by using higher spatial resolution in the $8 \times 8 \ \mathrm{km}^2$ box, while the bottom panels represent the  red box in Fig.~\ref{fig:16}. The overall amount of flavor conversion is comparable in the low and high resolution cases. However, due to the better spatial resolution,  small scale structures develop across a small patch  in the high-resolution run (see green patches in the top left panel of Fig.~\ref{fig:13}) in correspondence of the unstable regions found in the right panel of Fig.~\ref{fig:1a}. It is important to notice that the occurrence of a relatively larger conversion rate in a smaller spatial region in Fig.~\ref{fig:13}  does not lead to a spread of the flavor instability to nearby spatial bins. The overall flavor conversion rate averaged over a large area is thus unaffected by the presence of small scale structures.

\begin{figure}
 \centering
 \includegraphics[width=0.80\textwidth]{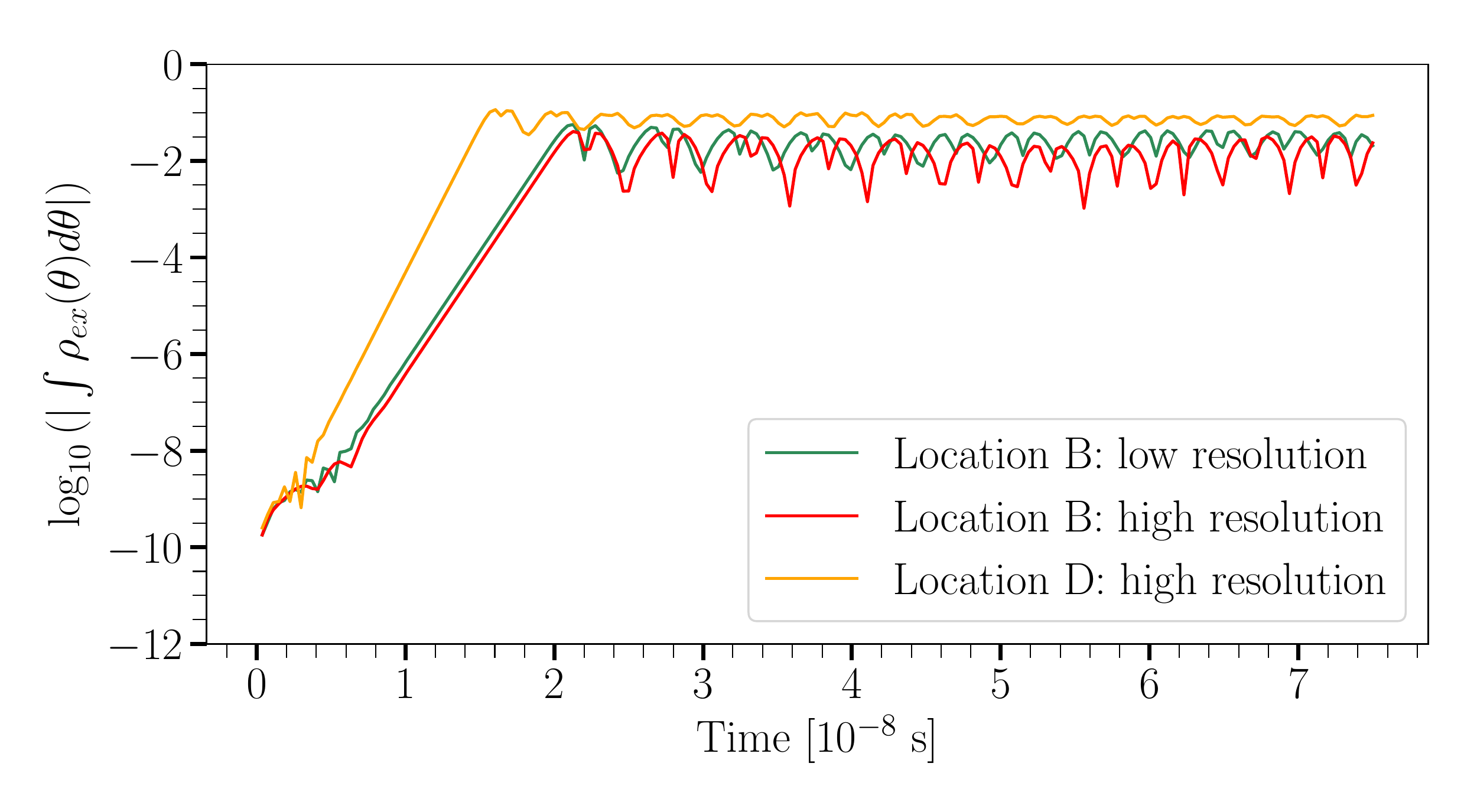}
 \caption{Same as in Fig.~\ref{fig:7a}, but here we compare the temporal evolution of the angle-integrated $\rho_{ex}$ elements of the low ($80 \times 80 \ \mathrm{km}^2$, green) and high ($8 \times 8 \ \mathrm{km}^2$, red and orange) spatial resolution grids for  points B and   D,  see also Fig.~\ref{fig:13}. The different spatial resolution negligibly affects the steady-state flavor configuration.
  }
 \label{fig:14}
 \end{figure}

Figure~\ref{fig:14} shows the time evolution of the angle-integrated $\rho_{ex}$ for point B in  the high resolution run  (in red) and in the low resolution run  (in green), and for point D in the high resolution run (in orange, see Fig.~\ref{fig:14}). Location D has been chosen as representative of the most unstable region in the top panels of Fig.~\ref{fig:13}. One can see that  the error in predicting the amount of flavor conversions in our low resolution runs is less than $1\%$ across the region inspected in  Fig.~\ref{fig:13} and up to $10\%$ for the small stripe showing the largest flavor conversions. Our findings are  not surprising  as the angular distributions for nearby bins are very similar to each other. 

Importantly, any spatial correlation between nearby spatial cells is averaged out by the advective term  in the non-linear regime, as discussed in Ref.~\cite{Shalgar:2019qwg}. Hence, the small localized region with slightly larger flavor conversions surrounding point D, in the high resolution run in Fig.~\ref{fig:13},  does not affect our overall conclusions. We stress that  the red box in Fig.~\ref{fig:16} corresponds to the region with the largest amount of flavor conversions, hence our spatial resolution allows to obtain even more accurate results for any remaining location above the remnant disk.
Given the negligible difference in the overall flavor outcome between the two runs with different spatial resolution, we choose to adopt the coarser grid throughout the paper since it allow us to explore a larger region above the remnant disk and better gauge the role of neutrinos in compact binary mergers.

Our results thus show the limitation of intuitive conclusions that can be drawn by relying on the linear stability analysis which imply a strong correlation between various spatial points. The collective nature of the flavor evolution is less manifest in the non-linear regime; this allows to perform numerical simulations  over a  coarser simulation grid than one may anticipate.

\end{appendix}

\newpage

\bibliographystyle{JHEP}
\bibliography{references}

\end{document}